%% file: FL_APR.tex
\documentclass{sig-alternate-05-2015}

\usepackage{latexsym}
\usepackage{amssymb}
\usepackage{ifthen}
\usepackage{graphicx}
\usepackage{microtype}
\usepackage{multirow}
\usepackage{cases}
\usepackage{bigstrut,multirow,rotating}
\usepackage{booktabs}
\usepackage{xcolor,stfloats}
\usepackage{listings}
\usepackage{balance}
\usepackage{hhline}
\usepackage{enumerate}
\usepackage[font=bf]{caption}
\usepackage{subcaption}
\usepackage{enumitem}
\usepackage{ntheorem}
\usepackage{xcolor,colortbl}
\usepackage{amsmath}
\usepackage{pifont}
\usepackage{url}
\usepackage{hyperref}
\usepackage{array}

\usepackage[linesnumbered,lined,boxed,commentsnumbered]{algorithm2e}
\SetAlFnt{\small}
\SetAlCapFnt{\small}
\SetAlCapNameFnt{\small}

\definecolor{grey}{rgb}{0.8,0.8,0.8}
\definecolor{darkgreen}{rgb}{0.17, 0.64, 0.37}

\hypersetup{
	colorlinks=true,      
	linkcolor=blue,       
	citecolor=blue,    
	filecolor=cyan,       
	urlcolor=black          
}

\let\origthelstnumber\thelstnumber
\makeatletter
\newcommand*\Suppressnumber{%
	\lst@AddToHook{OnNewLine}{%
		\let\thelstnumber\relax%
		\advance\c@lstnumber-\@ne\relax%
	}%
}

\newcommand*\Reactivatenumber[1]{%
	\lst@AddToHook{OnNewLine}{%
		\let\thelstnumber\origthelstnumber%
		\setcounter{lstnumber}{\numexpr#1-1\relax}%
	}%
}
\makeatother

\newcommand*{\all}{et al.}
\begin{document}
\setcopyright{acmcopyright}
%

\title{An Empirical Analysis of the Influence of Fault Space on \\ Search-Based Automated Program Repair}

\author{
	Ming Wen$^{1}$, Junjie Chen$^{2,3}$, Rongxin Wu$^{1}$, Dan Hao$^{2,3}$, Shing-Chi Cheung$^{1}$\\
	$^1$Department of Computer Science and Engineering \\
	The Hong Kong University of Science and Technology, Hong Kong, China \\
	\{mwenaa,wurongxin,scc\}@cse.ust.hk\\
	$^2$Key Laboratory of High Confidence Software Technologies (Peking University), MoE\\
	$^3$Institute of Software, EECS, Peking University, Beijing, 100871, China\\
	\{chenjunjie,haodan\}@pku.edu.cn\\
}
     
\maketitle

\input{abstract.tex}

\ccsdesc[500]{Software and its engineering~Automatic programming}
\ccsdesc[300]{Software and its engineering~Empirical software validation}

\printccsdesc

\keywords{Automated Program Repair, Fault Space, Empirical Study, Test Suite Augmentation}

\input{introduction.tex}
\input{APR.tex}

\input{experiment.tex}

\input{approach.tex}

\input{related.tex}
\input{conclusion.tex}

\balance
\bibliographystyle{abbrv}
\bibliography{bibliography}

\end{document}

%% file: abstract.tex
\begin{abstract}
Automated program repair (APR) has attracted great research attention, and various techniques have been proposed.
Search-based APR is one of the most important categories among these techniques. 
Existing researches focus on the design of effective mutation operators and searching algorithms to better find the correct patch.  
Despite various efforts, the effectiveness of these techniques are still limited by the search space explosion problem.
One of the key factors attribute to this problem is the quality of fault spaces as reported by existing studies.
This motivates us to study the importance of the fault space to the success of finding a correct patch. 
Our empirical study aims to answer three questions. Does the fault space significantly correlate with the performance of search-based APR? If so, are there any indicative measurements to approximate the accuracy of the fault space before applying expensive APR techniques? Are there any automatic methods that can improve the accuracy of the fault space? 
We observe that the accuracy of the fault space affects the effectiveness and efficiency of search-based APR techniques, e.g., 
the failure rate of GenProg could be as high as $60\%$ when the real fix location is ranked lower than 10 even though the correct patch is in the search space.
Besides, GenProg is able to find more correct patches and with fewer trials when given a fault space with a higher accuracy. 
We also find that the negative mutation coverage, which is designed in this study to measure the capability of a test suite to kill the mutants created on the statements executed by failing tests, is the most indicative measurement to estimate the efficiency of search-based APR.
Finally, we confirm that automated generated test cases can help improve the accuracy of fault spaces, and further improve the performance of search-based APR techniques.

\end{abstract}

%% file: introduction.tex
\section{Introduction}
A study conducted by Cambridge University shows that the global cost of general debugging is \$312 billion dollars annually and software developers spend 50\% of their time on fixing bugs \cite{britton2013reversible}.
The excessively high cost resulted from buggy software motivates the research on repairing bugs automatically.
Over the years, a large number of \textit{Automated Program Repair} (APR) techniques have been proposed \cite{kim2013automatic,le2012systematic,long2015staged,long2016automatic,mechtaev2015directfix,nguyen2013semfix,tan2015relifix,weimer2013leveraging,weimer2009automatically}.
Existing APR techniques can be broadly divided into two categories:
semantics-based APR (e.g., SemFix \cite{nguyen2013semfix}, Relifix \cite{tan2015relifix} and DirectFix \cite{mechtaev2015directfix}) and
search-based APR (e.g., GenProg \cite{le2012systematic,weimer2009automatically}, AE \cite{weimer2013leveraging}, PAR \cite{kim2013automatic}, SPR \cite{long2015staged} and Prophet \cite{long2016automatic}).
Semantics-based APR techniques synthesize a repair patch directly using semantic information via symbolic execution and constraint solving \cite{jin2011automated,mechtaev2015directfix,mechtaev2016angelix,nguyen2013semfix,wei2010automated}.
Search-based APR techniques, on the other hand, search within a huge population of candidate patches generated by applying mutation operators or a set of predefined repair templates.
Existing studies \cite{le2016towards,smith2015cure} have demonstrated that the two categories of APR techniques complement each other to some extent,
and they work in different settings, thus facing different challenges and limitations \cite{smith2015cure}.
Furthermore, search-based APR techniques are widely recognized to be able to fix a wide range of bugs and scalable to large programs without requiring extra specifications.
Therefore, in this paper, we focus on search-based APR techniques. 
Investigating semantics-based APR techniques is also of great value, but is outside the scope of this paper.

Search-based APR techniques work as follows. They produce a ranked list of the most suspicious code elements using \textit{fault localization} (FL) techniques \cite{qi2013using}. 
The ranked list is known as the \textit{fault space} \cite{le2013current, smith2015cure}.
The APR techniques then mutate these code elements to generate a large set of candidate patches. 
Next, they deploy searching heuristics \cite{kim2013automatic,le2012systematic,qi2014strength} to search among these candidates for a patch that passes all given test cases \cite{kim2013automatic,le2012systematic,perkins2009automatically,tan2015relifix,weimer2009automatically}.
These APR techniques generally suffer from the \textit{search space explosion} problem \cite{le2013current}. 
In addition, the correct patches are highly sparse in the large APR's search space \cite{long2016analysis}.
Le Goues et al. \cite{le2013current,smith2015cure} reported that the fault space critically determines APR's search space, and it is one of key hurdles that search-based APR must overcome to find correct patches \cite{weimer2013leveraging}.

A fault space $\mathcal{FS}$ prescribes the set of code elements to be mutated by APR techniques \cite{le2013current}. 
Each $\mathcal{FS} = \{\langle l_1, s_1\rangle,\langle l_2, s_2\rangle,\\...,\langle l_n, s_n\rangle\} (0 < s_i \leq 1\;and\;s_i \geq s_{i+1})$ is a ranked list of suspicious code elements, where $l_i$ represents the line number of a program and $s_i$ represents the corresponding suspicious score.
Search-based APR techniques generally assume a $\mathcal{FS}$ generated by fault localization techniques to generate candidate patches.
Listing \ref{motivatingExample} shows a real bug in project \textit{Apache Math}. We use it to illustrate that $\mathcal{FS}$ can affect a search-based APR's performance.
The original code misses a condition to check whether the current \texttt{Complex} and the targeting \texttt{rhs} are \texttt{null} before performing an addition operation.
Developers patched it by inserting the missing condition after line 152. 
To demonstrate the effect of $\mathcal{FS}$ on search-based APR, we fed three different fault spaces to GenProg \cite{weimer2009automatically,le2012systematic}. 
Table \ref{tab:fss} gives the top five elements of these three fault spaces, which rank the faulty line 152 at different positions.
GenProg is capable of generating a patch for all these fault spaces as shown in Listing \ref{motivatingExample}.

\begin{table} [t]
	\centering
	\small
	\caption{Descriptions of the Three Different Fault Spaces Generated by Different Test Suite}
	\label{tab:fss}
	\bgroup
	\setlength\tabcolsep{8pt}     
	\def\arraystretch{1.1}
	\begin{tabular}{r|c|c|c}
		\hline
		& $\mathcal{FS}$\#1 & $\mathcal{FS}$\#2 & $\mathcal{FS}$\#3 \\ \hline
		Top Elements & 152 : 0.904 &  319 : 0.929     &  997 : 1.000     \\ 
		 (\textit{line: susp})& 153 : 0.904 & 152 : 0.867      &  299 : 1.000     \\
		\multirow{3}{*}{}				& 319 : 0.873 &  153 : 0.867     &    308 : 1.000   \\
						& 997 : 0.715    &  997 : 0.666     &  152 : 1.000     \\
						& 308 : 0.660    &  308 : 0.565    &  153 : 1.000     \\ \hline 
		\multicolumn{4}{l}{\tiny{All the lines come from class \textbf{Complex.java}}, and line 152 is the oracle}  \\
	\end{tabular}
	\egroup
\vspace{-3mm}
\end{table}

\lstset{caption=Buggy class \textsc{Complex.java} of bug Math 53, stepnumber=1,firstnumber=151}
\begin{lstlisting}[language=Java,label=motivatingExample]
public Complex add(Complex rhs) {
  MathUtils.checkNotNull(rhs);*\Suppressnumber*
// Developer's Patch
+  if (isNaN || rhs.isNaN) {
+   return NaN;
+  }
// Patch 1:Generated by Fault Space #1
// +  if (isNaN || rhs.isNaN) {
// +    return org.apache.commons.math.complex.Complex.NaN;
// +  }*\Reactivatenumber{153}*
  return createComplex(real + rhs.getReal(),imaginary + rhs.getImaginary());*\Suppressnumber*
// Patch 2:Generated by Fault Space #2
// + return org.apache.commons.math.complex.Complex.NaN;*\Reactivatenumber{153}*
// - return createComplex(real + rhs.getReal(),imaginary + rhs.getImaginary());*\Reactivatenumber{154}*
}*\Suppressnumber*
// Patch 3:Generated by Fault Space #3
+ return org.apache.commons.math.complex.Complex.NaN;*\Reactivatenumber{997}*
- return new org.apache.commons.math.complex.Complex(realPart, imaginaryPart)*\Suppressnumber*
\end{lstlisting}

Patch 1 is a \textit{correct} solution, which is semantically the same as the developer's patch.
Patches 2 and 3 are \textit{plausible} solutions that pass all tests but semantically differ from the real patch.
From this example, the real fixing location is between line 152 and 153 in class \texttt{Complex}, and thus $\mathcal{FS}$\#1 has the highest accuracy.
Moreover, Patch 1, generating by $\mathcal{FS}$\#1, is the \textit{correct} solution. 
On the other hand, $\mathcal{FS}$\#3 possesses the lowest accuracy and produces the poorest patch, which differs from the developer's patch the most largely.
This indicates that the quality of fault spaces is correlated with the patches generated by search-based APR.
However, there is no study to systematically explore how does the fault space influence the performance of search-based APR techniques.
Are the capabilities of existing search-based APR techniques limited by the quality of fault spaces?
Is it possible for developers to approximate the quality of fault spaces before applying these expensive techniques?
Besides, can we improve the quality of fault spaces in practice?
Answering these questions helps tackle the search space explosion problem and facilitates researchers design new APR techniques in the future.


In this paper, we present a systematic empirical study that explores the influence of fault space on search-based APR techniques. 
First, we conducted controlled experiments to investigate the correlations between fault space accuracy and the performance (e.g., effectiveness and efficiency) of search-based APR techniques.
To conduct the experiments, we prepared plenty of fault spaces whose accuracy is uniformly distributed between 0 and 1 and then fed them to GenProg. 
In particular, all generated fault spaces contain the faulty code element. 
Our experimental results show that GenProg is capable of fixing more bugs correctly when fault spaces with high accuracy are fed.
Besides, GenProg is able to generate the correct patch faster (e.g., with less trials) when the accuracy of the fault space is higher.
Another interesting finding is that if the real fix location is ranked lower than 10 in the fault space, the failure rate of GenProg could be as high as 60\% even though the search space contains the correct patch.
These findings suggest that the capabilities of existing search-based APR techniques are largely limited by the quality of fault spaces. 
As such, generating fault spaces with high accuracy deserves more attention in the future when designing new search-based APR techniques. 

Second, we investigate how to approximate the accuracy of fault spaces before applying expensive search-based APR techniques.
When applying search-based APR technique, there are two major factors influencing the accuracy of fault spaces in practice: the spectrum-based FL approach used to compute the suspicious score \cite{abreu2009practical,qi2013using} and the test suite leveraged to generate the program spectrum.
The existing work \cite{abreu2009practical} has demonstrated that Ochiai performs better than the other techniques in general.
Therefore, in the second part of this study, we adopt Ochiai and focus on the other factor.
That is, we investigate whether the quality of test suites is correlated with the performance of search-based APR, and which measurement characterizes the correlation the most strongly.
Our experimental results show that all the different coverage criteria (i.e., line, branch and mutation coverage) are highly correlated with the effectiveness and efficiency of search-based APR techniques.
Besides, the \textit{negative mutation coverage}, which is designed in this study to measure the capabilities of a test suite by calculating the proportion of killed mutants that are generated on the statements along the failing execution trace, correlates with the performance the most strongly.
These findings indicate that we can approximate the performance of search-based APR techniques before applying them by measuring the coverages of the corresponding test suite, and the negative mutation coverage is the most indicative information.

Third, we augmented the test suite by leveraging automated test generation tools (e.g., Randoop and Evosuite) in order to generate more accurate fault spaces based on our findings in the second part.
Experimental results confirm the potential that feeding the fault spaces generated by leveraging the automatically generated test cases help improve the efficiency of existing search-based APR techniques.

The rest of this paper is structured as follows.
Section 2 introduces the background of search-based APR techniques.
The setup of all the experiments is introduced in Section 3.
In Section 4, we conduct a series of experiments to analysis the influence of fault space on search-based APR techniques.
Section 5 discusses the related works and Section 6 concludes this work.

%% file: APR.tex
\section{Search-Based APR}
\label{APR}

Search-based APR, which may also be referred as \textit{generate and validate} APR techniques in some literature \cite{long2016analysis,qi2015analysis,smith2015cure}, is able to fix a wide range of bugs and scalable to real programs in large scale \cite{kim2013automatic,le2012systematic,perkins2009automatically,tan2015relifix,weimer2009automatically}.
A typical APR process consists of four steps: \textit{fault localization}, \textit{patch gneration}, \textit{prioritization \& searching} and \textit{validation}.
Step 1 produces the fault space ($\mathcal{FS}$), i.e, the suspicious code elements related to the fault needing to be fixed.
Different APR techniques leverage different FL techniques to generate the fault spaces, such as Tarantula \cite{nguyen2013semfix}, Ochiai \cite{ke2015repairing,tan2015relifix,martinez2016astor,xuan2016nopol} and Jaccard \cite{mechtaev2016angelix}. 
The performance of these different FL techniques in APR have been recently studied \cite{qi2013using,assiri2016fault}.
Step 2 defines the \textit{fix space}, i.e, the operators can be applied to each of the code element.
For example, GenProg \cite{weimer2009automatically,le2012systematic} leverages two types of operators: mutation operator (i.e., code insertion, removal or replacement) and crossover operator to generate candidate patches.
PAR \cite{kim2013automatic} defines the operators by templates learned from human-written patches.
For some operators like \textit{insertion} and \textit{replacement}, code ingredients are required in order to generate a candidate patch.
Therefore, existing APR techniques define the \textit{ingredients space}, i.e, the space to search for source ingredients.
The ingredients can be searched within the same application \cite{le2012systematic,long2015staged,qi2014strength,weimer2013leveraging} or across other applications \cite{sidiroglou2015automatic}.
Step 3 defines the algorithm adopted to search the correct patch among the candidate patches.
Searching heuristics like genetic programming \cite{kim2013automatic,le2012systematic} and random search \cite{qi2014strength} have been used by existing studies.
For genetic programming, the searching procedure is guided by the fitness score of each candidate, which is computed using the number of failing and passing test cases by running the candidate against the provided test suite.
The generations keep iterating until a valid patch passing all test cases is obtained or some limits arrives (i.e., time budget or the max number of generations).
The last step is to validate whether a generated patch is correct or not by running it against the provided test suite.

The fault space, fix space and ingredients space described above jointly attribute to the search space explosion problem, which still remains to be one of the major open challenges for search-based APR techniques \cite{le2013current}. 
To address this problem, a lot of efforts have been mostly made to design more accurate fix space in the second step \cite{kim2013automatic,le2016history,long2015staged,weimer2013leveraging}, and to search the search space more efficiently for a correct patch in the third step \cite{long2016automatic,qi2014strength}.
However, no systematic study has yet been made to characterize the influence of the fault space produced in the first step on the efficiency and effectiveness of APR techniques.
This study aims to bridge this gap.

%% file: experiment.tex
\section{Experimental Setup}

We select GenProg \cite{le2012systematic,weimer2009automatically} in this study for the following reasons:
First, it is one of most representative search-based APR techniques and has been widely used and evaluated in existing literature \cite{qi2013efficient,weimer2013leveraging,long2015staged,xuan2016nopol,martinez2016astor}.
Second, evidence was shown that GenProg is capable of generating patches sufficiently often to enable empirical studies \cite{smith2015cure}.
For such a reason, GenProg has been widely selected as the subject for conducting empirical studies \cite{assiri2016fault,qi2014strength,qi2013using,qi2015analysis,smith2015cure}.
Third, a Java version of GenProg is publicly available \cite{martinez2016astor} to facilitate our controlled experiments and also for the replications of the results. 
In this study, we modify GenProg based on the Java implementation \cite{martinez2016astor} to make it capable of accepting any fault spaces as input to generate candidate patches in order to conduct the controlled experiments.
We use Ochiai to generate fault spaces on different test suites since the original implementation adopts it \cite{martinez2016astor}. 
Besides, Ochiai is reported to achieve better performance than the other techniques \cite{abreu2009practical}.
To determine the size of the fault spaces, different APR techniques adopt different heuristics.
For example, GenProg and PAR set the threshold as 0.1, which means they keep only those statements whose suspiciousness is over 0.1 in the fault space.
In this study, we adopt another heuristic that is also widely used by existing techniques \cite{long2015staged,long2016analysis,long2016automatic}, which is to select the top N statements from the results produced by FL techniques.
And we set N to 100 similar to the existing study \cite{long2016analysis}.
Another setting to enable the successful run of GenProg is the scope in which the statements are manipulated to find a patch.
We set it to the package where the bug occurred similar to existing studies \cite{martinez2016astor,martinez2016automatic}, and we call it the \textit{target package}.
In particular, various coverage criteria are also measured on the target package in the later of this studies.
The time budget is set to be one hour for each run. 
A repair operation is regarded as ``failed'' if no patches are generated within the time budget. 

\subsection{The Subject Programs}

In our study, we use the Defects4J benchmark \cite{just2014defects4j}, which consists of 357 real bugs extracted from five Java open source projects.
The benchmark was built to facilitate controlled experiments in software debugging and testing researches \cite{just2014defects4j}, and has been widely adopted by recent studies \cite{le2016history,martinez2016automatic,martinez2016astor,shamshiri2015automatically,xuan2016nopol}.
Following the existing work on automated program repair\cite{martinez2016automatic,shamshiri2015automatically}, we exclude the Closure Compiler project, because it does not use standard JUnit test cases.
That is to say, we use four projects in the Defects4J benchmark as subjects in our study.
There are altogether 224 program versions in these four projects. Table \ref{tab:dataset} presents the basic information of subjects.
\begin{table} [t]
	\centering
	\caption{Subjects}
	\label{tab:dataset}
	\bgroup
	\setlength\tabcolsep{2.8pt}     
	\def\arraystretch{1.3}
	\begin{tabular} {lcccc}
		\hline
		Subject &  \#Bugs & KLOC & Test KLOC & \#Test Cases  \\ \hline
		Commons Lang & 65 & 22 & 6 & 2,245  \\ \hline
		JFreeChart & 26 & 96 & 50 & 2,205  \\ \hline
		Commons Math & 106 & 85 & 19 & 3,602 \\ \hline
		Joda-Time & 27 & 28 & 53 & 4,130 \\ \hline
		Total & 224 & 231 & 128 & 12,182\\ \hline
	\end{tabular}
	\egroup
\vspace{-3mm}
\end{table}

\subsection{Measurements}

GenProg \cite{le2012systematic,weimer2009automatically} is reported to be able to fix 55 out of 105 real bugs on their dataset by validating if they pass all given tests.
However, later studies \cite{qi2015analysis} reveal that a patch may not be correct even though it passes all given test cases. This is because the given test cases in these subjects are an imperfect validation of program correctness. Validation using these test cases cannot discriminate correct patches from \textit{overfitting} ones \cite{smith2015cure}.
The term \textit{plausible} has been widely used to denote those patches that pass all tests in the given test suite \cite{long2015staged,long2016analysis,qi2015analysis}.
Moreover, patches are considered as \textit{correct} if they pass all given test cases and are semantically the same as the patches submitted by developers.
The effectiveness of APR is measured by the number of \textit{plausible} and \textit{correct} patches.
Table \ref{tab:genprog} summarizes the effectiveness of GenProg on the Defects4J benchmark as reported by Martinez \all \cite{martinez2016automatic}.
\begin{table} [ht]
	\centering
	\caption{Effectiveness of GenProg on Defects4J}
	\label{tab:genprog}
	\bgroup
	\setlength\tabcolsep{2.8pt}     
	\def\arraystretch{1.3}
\begin{tabular}{rccccc}
	\hline
	& Math & Chart & Time & Lang & Total\\ \hline
	plausible & 18   & 7     & 2    & 0 & 27   \\
	correct   & 5    & 0     & 0    & 0 & 5  \\ \hline
\end{tabular}
\egroup
\vspace{-2mm}
\end{table}

Even if an APR technique is able to generate a correct patch, it still may not find it within an affordable time budget since correct patches are highly sparse in the search space \cite{long2016analysis,long2016automatic}.
Existing studies show that the efficiency of APR techniques is greatly compromised by the search space explosion problem \cite{long2016analysis}.
In this study, we measure efficiency using the \textit{number of patches searched} (NPS) before a solution is found.
This measurement has been widely adopted by the existing studies \cite{qi2013using,assiri2016fault,qi2015analysis,long2016automatic}.

\subsection{Correlation Analysis}

Our study also investigates by means of correlation analysis the relations between APR's efficiency and various factors, including fault space accuracy and test coverage.
Here, we use Spearman Rank Correlation to quantitatively study the relations since it makes no assumptions about the distributions of the variables and the relations can be non-linear.
Spearman values of $+1$ and $-1$ indicate strong positive or negative correlation, while $0$ indicates no correlation.
We also apply significance tests to examine the confidence level of the correlations.

%% file: approach.tex
\section{Empirical Evaluation}
Our empirical study consists of three parts.
Section \ref{sec:faultspace} studies the correlation between the accuracy of fault space and the performance of APR.
Second, we investigate the correlations between different coverage criteria and the performance of APR in Section \ref{sec:coverage}, since test suite quality is one of the major factor which affects the accuracy of the generated fault spaces in practice.
Note that we only control the quality of test suite to generate fault spaces, and still use the full set of tests for validation in order to control the variable and avoid the overfitting problem \cite{smith2015cure}.
Third, in Section \ref{sec:automatedTest}, we conduct studies to see if automated generated test cases can boost the performance of search-based APR.

\subsection{Accuracy of Fault Space}
\label{sec:faultspace}

Recently, many search-based APR techniques have been proposed \cite{kim2013automatic,le2012systematic,perkins2009automatically,tan2015relifix,weimer2009automatically}.
However, whether the capabilities of these techniques are limited by the repair algorithms or the fault space produced by the fault localization techniques remains unknown.
Revealing it can help researchers design new APR techniques in the future.
Driven by this, we propose the following research question:
\vspace{2mm}

\noindent\fbox{
	\parbox{0.95\linewidth}{
		\textbf{Research Question 1:} How does the accuracy of fault space influence the performance of search-based APR?
		
	}
}

\vspace{2mm}
We answer this question by investigating the following two aspects: 
can we repair more bugs (both plausibly and correctly) by improving the accuracy of fault spaces? (in terms of effectiveness)
and can we reduce the number of the patches searched before finding a patch by improving the accuracy of fault spaces? (in terms of efficiency).
Before answering these questions, we first introduce how to measure the accuracy of a fault space.

\subsubsection{Measuring the Accuracy of Fault Space}

We leverage the real fixing patched submitted by developers to extract the oracles of fix locations. 
We denote the set of real fix locations as $\mathcal{RF}$, and $l\in \mathcal{RF}$ if statement $l$ is modified by the real fixing patches.
The fault space $\mathcal{FS}$ contains a ranked list of suspicious lines.
We measure the accuracy of a $\mathcal{FS}$ by comparing it with the oracle $\mathcal{RF}$ to see how the real fix locations are ranked in the fault space.
Specifically, we leverage the widely-used metrics \textit{MAP} and \textit{MRR} in fault localization research \cite{abreu2009practical,abreu2007accuracy,wen2016locus} to quantify the accuracy of fault spaces.

\textbf{MRR:} \textit{Mean Reciprocal Rank} \cite{voorhees1999trec} is the average of the reciprocal ranks of  a set of queries.
This is a statistic for evaluating a process that produces a list of possible responses to a query. 
The reciprocal rank of a query is the multiplicative inverse of the rank of the first relevant answer found. The formal definition of MRR in the context of this is as follows: 
	\vspace{-2mm}
	\begin{equation}
	\textit{MRR} = \frac{1}{|Q|} \sum_{i = 1} ^{|Q|}\frac{1}{\textit{FR}_i}
	\vspace{-2mm}
	\end{equation}
	$FR_i$ represents the rank of the first real fix location $l$ ($l \in \mathcal{RF}$) found in a $\mathcal{FS}$. 
This metric is used to evaluate the ability to locate the first relevant code element for a bug.

\textbf{MAP:}	
\textit{Mean Average Precision} provides a single value measuring the quality of a ranked list, it takes all the relevant answers into consideration with their ranks for a single query. 
	The average precision is computed as:
	\vspace{-2mm}
	\begin{equation}
	AvgP = \frac{1}{N}\sum_{k=1}^n rel(k) P(k)
	\vspace{-1mm}
	\end{equation}
	where $k$ is the rank in the sequence of the retrieved answers, $n$ is the size of $\mathcal{FS}$ while $N$ denotes the size of $\mathcal{RF}$. In this formula, $rel(k)$ is an indicator function equaling one if the item at rank $k$ belongs to $\mathcal{RF}$, and zero otherwise. 
	$P(k)$ is the precision at the given cut-off rank $k$, which is computed as $P(k) = \frac{N_r}{k}$.
	where $N_r$ is the number of real fix locations in the top $k$ of $\mathcal{FS}$.
	MAP is the mean of the average precision, which is computed as:	
	\vspace{-1mm}
	\begin{equation}
	\textit{MAP} = \frac{\sum_{i=1}^{|Q|}AvgP(i)}{|Q|}
	\vspace{-1mm}
	\end{equation}
This metric is used to evaluate how well are all real fix locations in $\mathcal{RF}$ being ranked in the $\mathcal{FS}$. 
For both metrics, a larger value indicates a fault space with higher accuracy.
\newline

\begin{figure}[h!]
	\centering
	\includegraphics[width=1.0\linewidth]{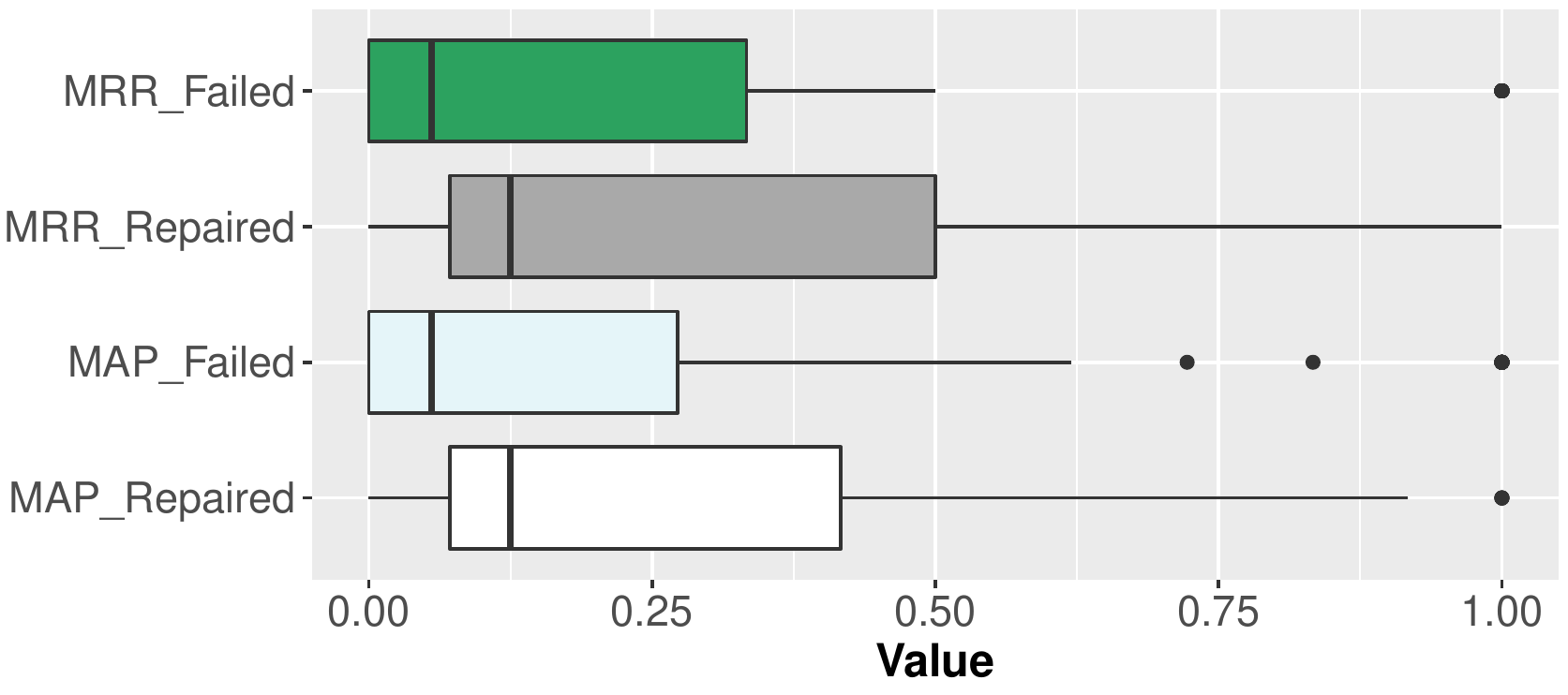}
	\caption{Statistics of the Accuracy of Fault Spaces} 
	\label{fig:MAP_MRR}
\end{figure}

Figure \ref{fig:MAP_MRR} shows the statistics of the MAP and MRR values of the fault spaces generated by the provided test suite using Ochiai for all the bugs.
We classified them into two groups, those can be repaired by GenProg (e.g., marked with MAP\_Repaired) and those can not (e.g., marked with MAP\_Failed).
The Mann-Whitney U-Test \cite{mann1947test} indicates the values of MAP\_Repaired is significantly larger than those of MAP\_Failed (p.value = 0.011).
Similar results were found on MRR (p.value = 0.009). 
This reveals that a bug is more likely to be repaired when the accuracy of the generated fault space is higher.


Instead of using spectrum-based FL techniques to generate the fault space, we generate a large number of fault spaces whose accuracy is uniformly distributed between 0 and 1 in order to conduct the controlled experiments.
Specifically, we set 10 target ranges, from (0,0.1], (0.1,0.2] to (0.9,1] for both MAP and MRR.
To generate a fault space $\mathcal{FS}$, we randomly select $N$ statements from those statements executed by any of the failing test cases and randomly rank these $N$ statements.
We then calculate the MAP and MRR of this $\mathcal{FS}$.
Please be noted that it is possible that no fault space can be generated for certain ranges.

We first use the fault spaces generated in the last range, whose accuracies are over 0.9, to see if more bugs can be repaired by providing fault spaces with such high qualities.
We then investigate if the accuracy of fault space correlates with the efficiency of search-based APR.
In order to conduct correlation analysis, we apply the technique on each bug 200 times, providing fault spaces with various accuracies each time, and the time budget is set to 1 hour for each run.
It is necessary to possess enough number of fault spaces with different ranges of accuracies.
Therefore, we prepare 20 fault spaces for each range. 
We feed each generated fault space to search-based APR, and then analyze the results of all runs to investigate the correlations between the accuracy of fault spaces and the performance of search-based APR.


\subsubsection{Repairing More Bugs}
\label{RQ1_effectiveness}
\begin{table} [t]
	\centering
	\small
	\caption{The Number of Bugs that can be Fixed by Improving the Accuracy of Fault Space}
	\label{tab:numfixedbugs}
	\bgroup
	\setlength\tabcolsep{3pt}     
	\def\arraystretch{1.1}
	\begin{tabular}{lcc|l|l}
		\hline
		Project & GenProg & GP$^\natural$ & New Plausible & New Correct \\ \hline
		Math    & 18   & 21    & \#20, \#44, \#56  & - \\ \hline
		Chart   & 7  & 9            & \#12, \#18  & \#1, \#12    \\ \hline
		Time   & 2   & 2             & -  & -   \\ \hline
		Lang   & 0   & 5             & \#7 \#22 \#24 \#38 \#39   & \#24 \#38   \\ \hline
		Total   & 27  & 37           & -   & -   \\ \hline
		\multicolumn{5}{l}{\scriptsize{GP$^\natural$ denotes GenProg with $\mathcal{FS}$ whose accuracy is over 0.9}}
	\end{tabular}
	\egroup
\vspace{-5mm}
\end{table}
First, we are interested in whether more bugs can be fixed either plausibly or correctly by feeding fault spaces with high accuracies, and the results are shown in Table \ref{tab:numfixedbugs}.
We find that 10 new bugs, which is 37\% more bugs, can be fixed plausibly by increasing the accuracy of fault spaces.
The corresponding bug id has been displayed in column \textit{New Plausible}.
We also investigate all the plausible patches manually, and find that for 4 new bugs, which are Chart\#1, Chart\#12, Lang\#24 and Lang\#28, GenProg actually produces correct solutions when better fault spaces are provided, and these results were not reported by the latest study \cite{martinez2016automatic}.
Listing \ref{chart1}, Listing \ref{chart12} and Listing \ref{lang38} shows the patches generated by GenProg and the corresponding patches submitted by developers.
A condition negation is able to fix Chart 1.
For Chart 12, some operations are required when the variable \texttt{dataset} is \texttt{null}.
Lang 38\footnote{https://issues.apache.org/jira/browse/LANG-538} relates to a ignorance of API preconditions, calling \texttt{Calenar.getTime()} is able to fix the problem.
All these patches can be generated by the operators that GenProg leverage, the key problem lies in whether these correct locations can be fed to GenProg or not.
We checked the accuracy of these four bugs, for both MAP and MRR, the values are 0.125, 0.063, 0.0 and 0.0, which shows that these real fix locations are ranked at very low positions in the fault space or even do not exist in the fault space at all.
These indicate that the effectiveness of existing search-based APR techniques is greatly compromised by the quality of fault spaces.

\begin{figure}[h]
	\centering
	\includegraphics[width=1.0\linewidth]{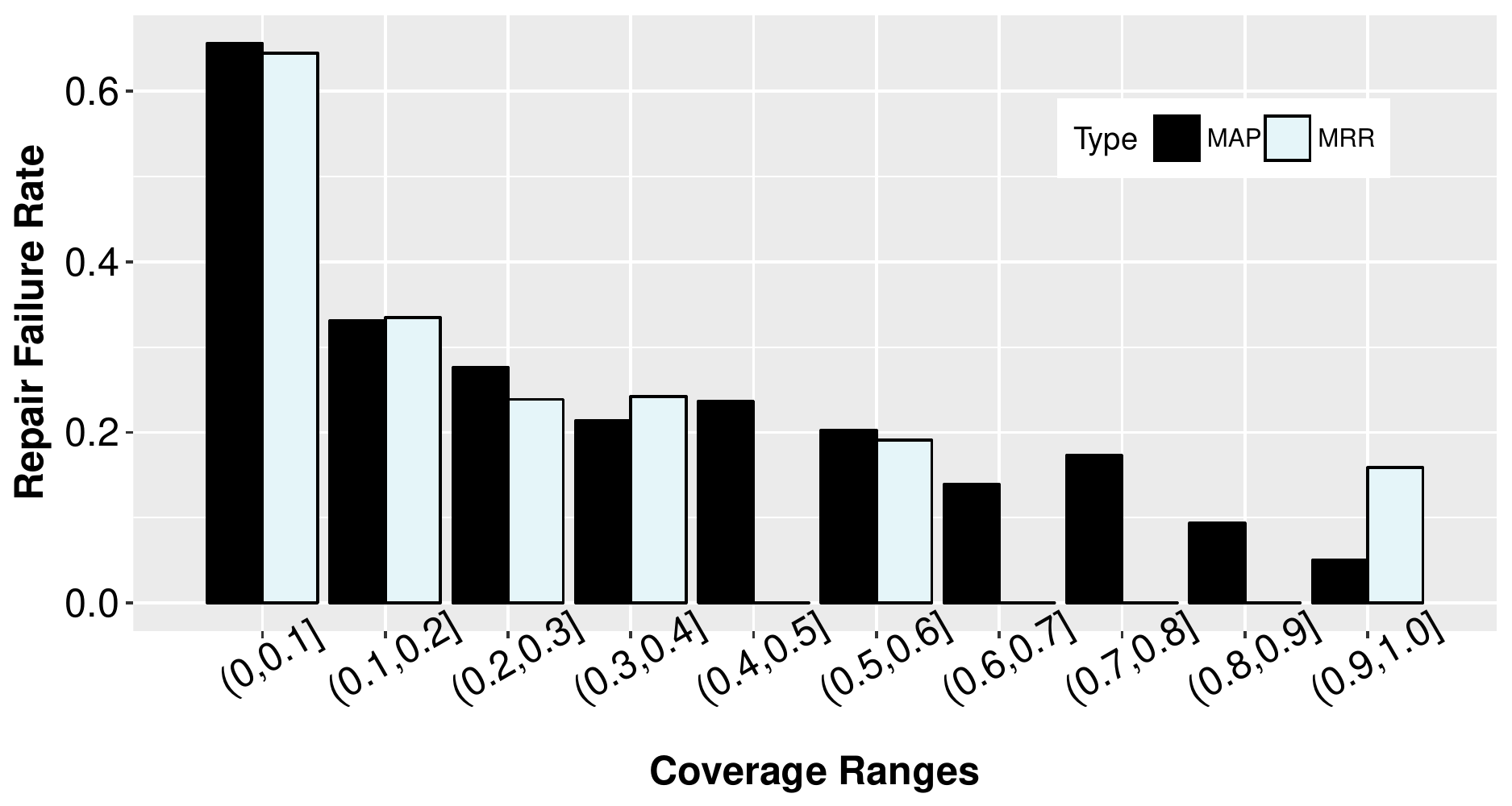}
	\caption{The Failure Rate of Different Fault Space Accuracy Values} 
	\label{fig:successRate}
\vspace{-3mm}
\end{figure}

For those bugs whose plausible solutions can be found for any of the 200 runs, which means GenProg is capable of generating a solution in the search space, we calculate the repairing failure rate for each range.
A repair run is regarded as failed if no patch is generated within the time budget.
We observe that the failure rate of GenProg is highly correlated with the accuracy of fault spaces.
Figure \ref{fig:successRate} shows the failure rates under the 10 ranges of fault space accuracies.
The results show that if the real fix location is ranked after 10, in which the value of MAP and MRR is smaller than 0.1, the failure rate could be as high as 60\% even though that the solution is in the search space.

\lstset{caption=Correct Patch of Chart 1, stepnumber=1,firstnumber=1797,xleftmargin=.045\textwidth}
\begin{lstlisting}[language=Java,label=chart1]
// developer's patch *\Reactivatenumber{1797}*
-  if (dataset != null) {*\Suppressnumber*
+  if (dataset == null) {
// patch generated by GenProg
// +  if (dataset == null) {*\Reactivatenumber{1798}*  
     return result;*\Reactivatenumber{1799}*
   }*\Reactivatenumber{871}*
\end{lstlisting}
\vspace{-5mm}
\lstset{caption=Correct Patch of Lang 38, stepnumber=1,firstnumber=1797,xleftmargin=.04\textwidth}
\begin{lstlisting}[language=Java,label=lang38]
if (mTimeZoneForced) {*\Suppressnumber*
// developer's patch 
+  calendar.getTime();
// patch generated by GenProg
// +  int value = calendar.getTime();*\Reactivatenumber{872}*
  calendar = (Calendar) calendar.clone();*\Reactivatenumber{873}*
  calendar.setTimeZone(mTimeZone);*\Reactivatenumber{874}*
}*\Reactivatenumber{144}*
\end{lstlisting}
\vspace{-4mm}
\lstset{caption=Correct Patch of Chart 12, stepnumber=1,xleftmargin=.04\textwidth}
\begin{lstlisting}[language=Java,label=chart12]
public MultiplePiePlot(CD dataset) {*\Reactivatenumber{145}*
 this.dataset = dataset;*\Suppressnumber*
// developer's patch*\Reactivatenumber{146}*
+ setDataset(dataset);*\Suppressnumber*
// patch generated by GenProg
// + if (dataset != null) {
// +  setDatasetGroup(dataset.getGroup());
// +  dataset.addChangeListener(this);
// + }
*\Reactivatenumber{175}*
public void setDataset(CD dataset) {*\Reactivatenumber{176}*
 if (dataset != null) {*\Reactivatenumber{175}*
  setDatasetGroup(dataset.getGroup());*\Reactivatenumber{177}*
  dataset.addChangeListener(this);*\Reactivatenumber{178}*
 }*\Reactivatenumber{179}*
}
\end{lstlisting}
\vspace{-2mm}

\subsubsection{Reducing the Number of Patches Searched}
\label{RQ1_effeciency}

For those bugs whose plausible solutions can be generated for at least 20 times during the 200 runs, we apply Spearman Rank Correlation to study the correlations between the fault space accuracies and the number of patches searched before the solution is found, and Table \ref{tab:correlation} shows the correlations and significant level.
We do not display the results for those bugs with less than 20 successful runs since it may not be sufficient to achieve statistically significant results if the number of samples is less than 20 \cite{bonett2000sample,smith2015cure}.

\begin{table} [ht]
	\centering
	\caption{Correlations between the Accuracy of Fault Space and the Repair Efficiency}
	\label{tab:correlation}
	\bgroup
	\setlength\tabcolsep{5.6pt}     
	\def\arraystretch{1}
	\small
	\begin{tabular}{lc|lc|lc}
		\hline
		\multirow{2}{*}{Project} & \multirow{2}{*}{Bug Id} & \multicolumn{2}{c|}{MAP} & \multicolumn{2}{c}{MRR} \\
		&                         &  Cor    & P.Value      &  Cor    & P.Value      \\ \hline
		Math & 5$^\surd$  & -0.944$^\star$ & 4.22E-49 & -0.944$^\star$ & 4.22E-49    \\
		Math & 50$^\surd$ & -0.562$^\star$ & 2.36E-07 & -0.624$^\star$ & 3.59E-09    \\
		Math & 53$^\surd$ & -0.668$^\star$ & 6.82E-27 & -0.669$^\star$ & 6.82E-27    \\
		Math & 70$^\surd$ & -0.912$^\star$ & 1.17E-41 & -0.912$^\star$ & 1.17E-41    \\
		Math & 73$^\surd$ & -0.363$^\star$ & 2.03E-04 & -0.462$^\star$ & 1.32E-06    \\
		Chart & 1$^\surd$ & -0.660$^\star$ & 0.002 & -0.660$^\star$ & 0.002 \\ 
		Chart & 12$^\surd$ & -0.237$^\star$ & 0.018 & -0.237$^\star$ & 0.018 \\ 
		Lang & 24$^\surd$ & -0.520$^\star$ & 0.001 & -0.520$^\star$ & 0.001 \\
		Lang & 38$^\surd$ & -0.449$^\star$ & 0.000 & -0.449$^\star$ & 0.000 \\ \hline
		Math & 28 & -0.491$^\star$ & 2.18E-07 & -0.666$^\star$ & 3.83E-14    \\
		Math & 44 & -0.537$^\star$ & 8.42E-09 & -0.514$^\star$ & 4.55E-08    \\
		Math & 80 & -0.098 & 0.482    & -0.098 & 0.482       \\
		Math & 81 & -0.342$^\star$ & 0.002    & -0.574$^\star$ & 6.04E-08    \\
		Math & 84 & -0.624$^\star$ & 3.05E-10 & -0.560$^\star$ & 3.74E-08    \\
		Math & 85 & -0.214 & 0.067    & -0.214 & 0.067       \\
		Math & 95 & -0.494$^\star$ & 1.72E-07 & -0.456$^\star$ & 1.91E-06  \\ 
		Lang & 7 & -0.217 & 0.082 & -0.249$^\star$ & 0.044 \\
		Lang & 22 & -0.404$^\star$ & 0.001 & -0.449$^\star$ & 8.75E-05 \\ \hline
		\multicolumn{6}{l}{$\surd$ means the patch is correct, and $\star$ means the correlations} \\
		\multicolumn{6}{l}{are significant ($p < 0.05$).}       
	\end{tabular}
	\egroup
	\vspace{-5mm}
\end{table}

\begin{figure*} [t]

	\begin{subfigure}{.195\textwidth}
		\includegraphics[width=1.0\linewidth]{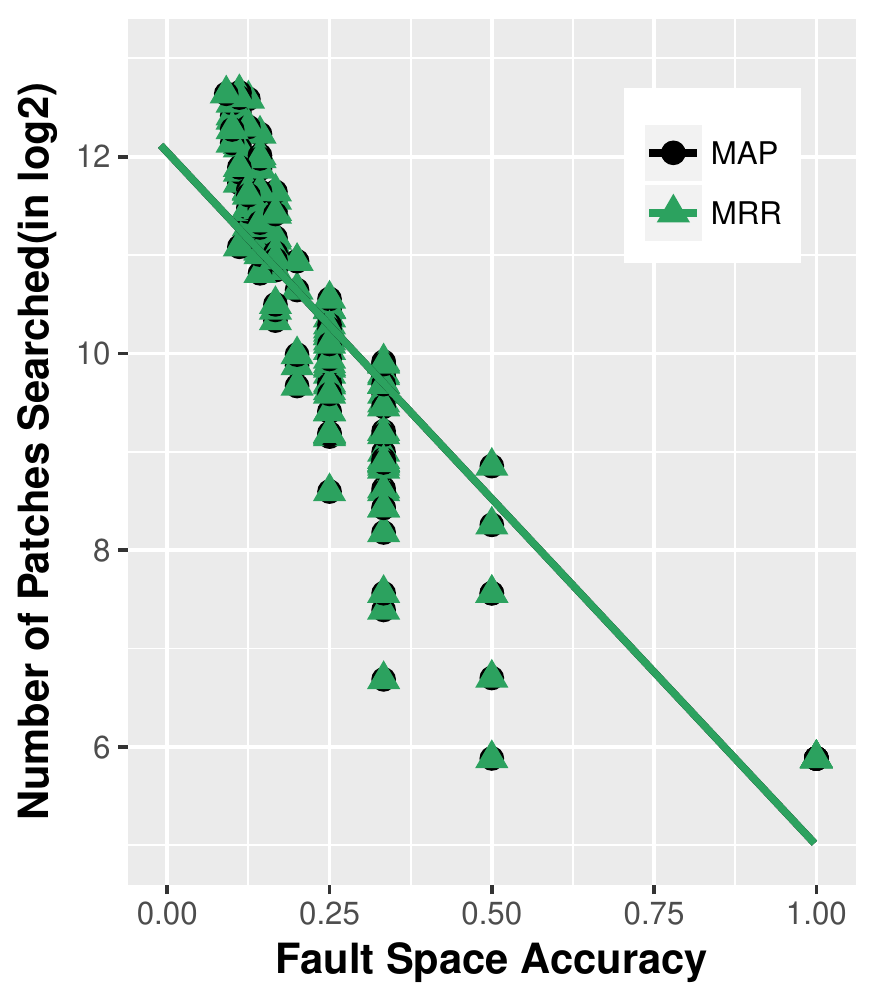}
		\caption{Math 5}
	\end{subfigure} 
	\begin{subfigure}{.195\textwidth}
		\includegraphics[width=1.0\linewidth]{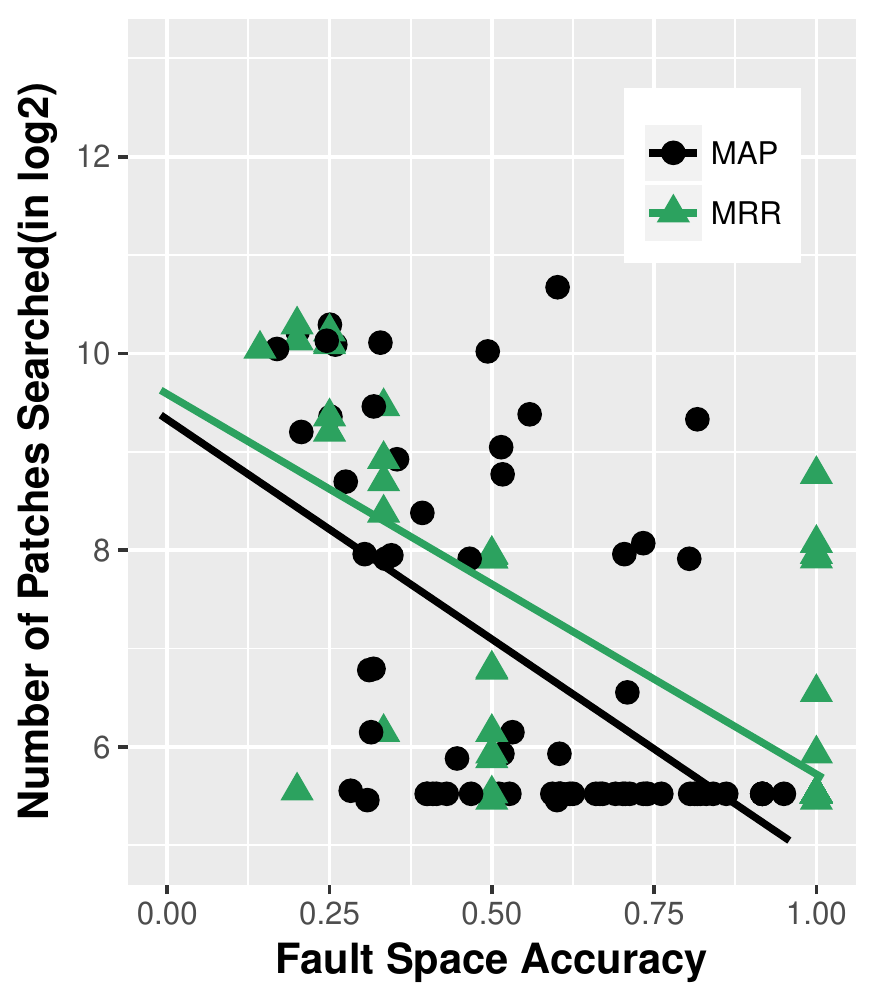}
		\caption{Math 50}
	\end{subfigure}
	\begin{subfigure}{.195\textwidth}
		\includegraphics[width=1.0\linewidth]{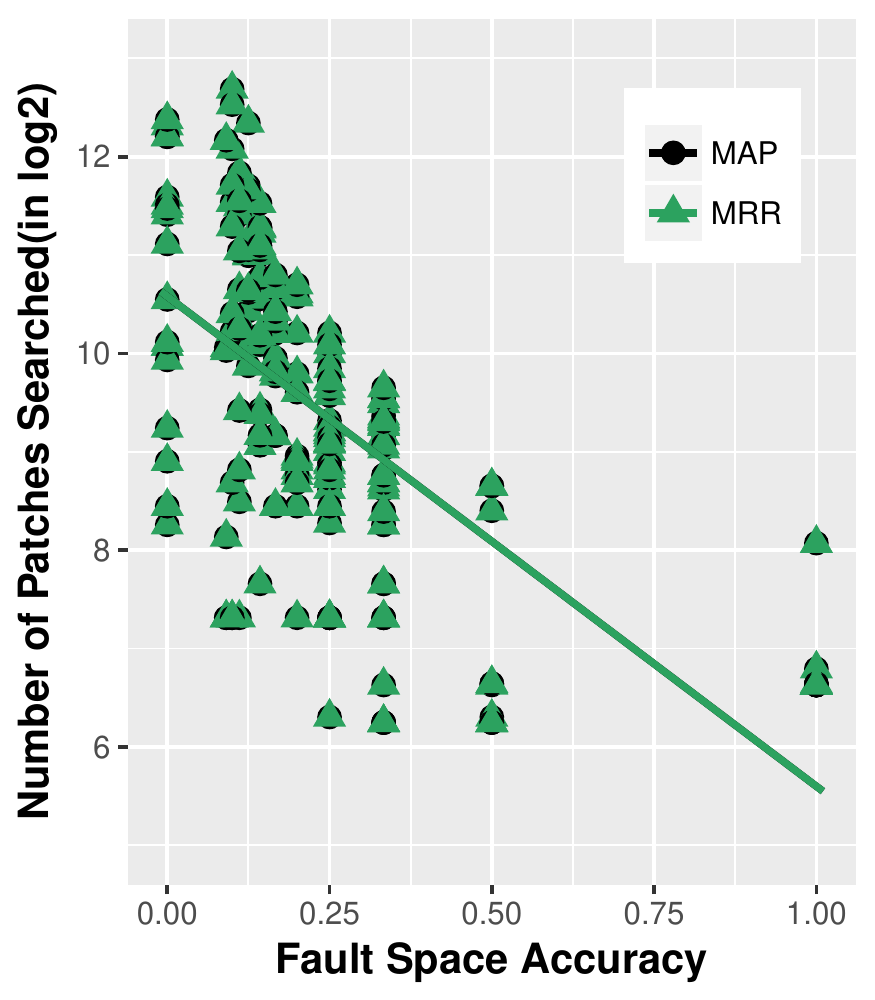}
		\caption{Math 53}
	\end{subfigure}
	\begin{subfigure}{.195\textwidth}
		\includegraphics[width=1.0\linewidth]{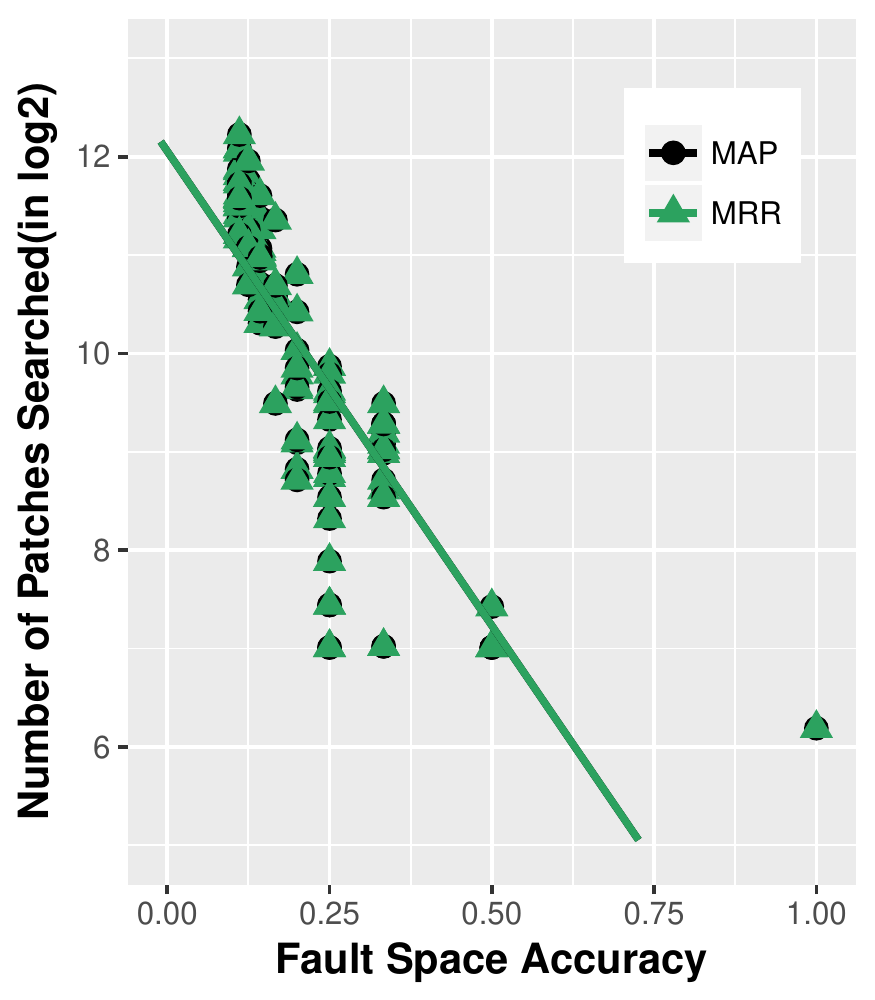}
		\caption{Math 70}
	\end{subfigure}
	\begin{subfigure}{.195\textwidth}
	\includegraphics[width=1.0\linewidth]{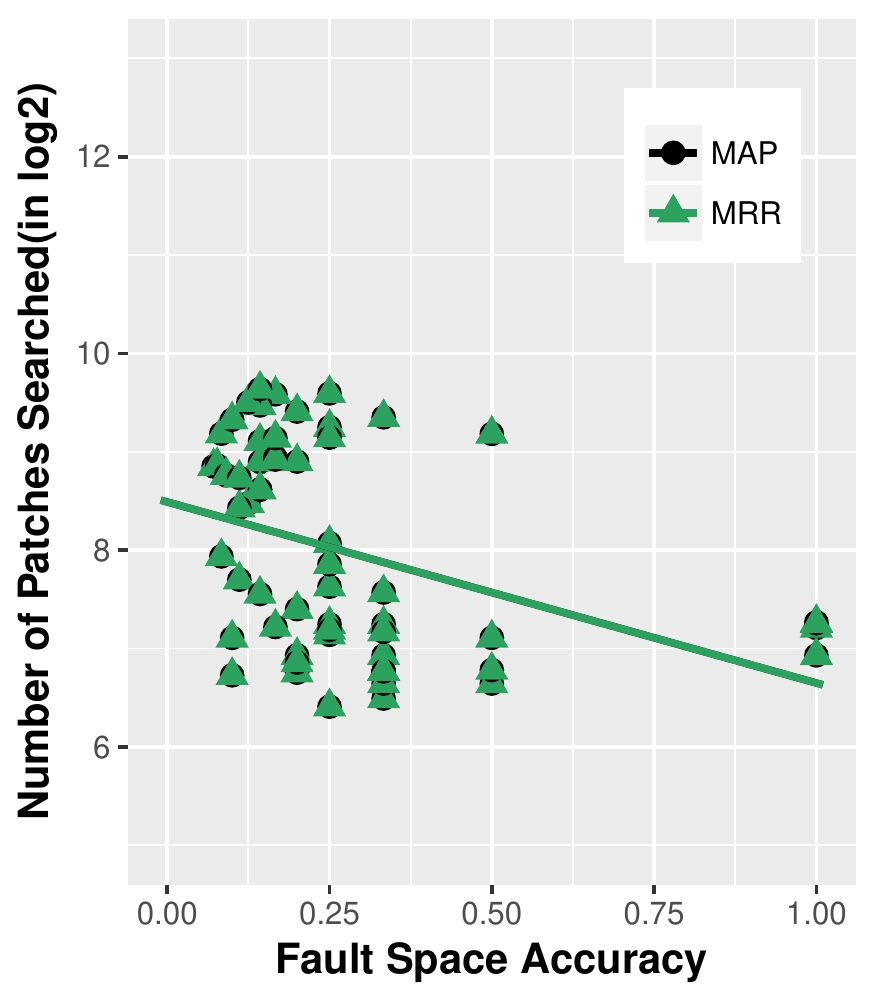}
	\caption{Lang 38}
	\end{subfigure}
	\caption{The Correlation between the Accuracy of Fault Space and the Search Space Size}
	\label{fig:searchSpace}
\vspace{-3mm}
\end{figure*}

The results show that the accuracy of fault space, for both MAP and MRR, is negatively correlated with the number of patches searched before a solution is generated.
For most of the cases, 15 out of 18 for MAP, and 16 out of 18 for MRR, the correlations are significant ($p < 0.05$).
It reveals that GenProg is able to find a patch with less trials if fault spaces with higher accuracies are provided.
We sample five bugs which can be correctly repaired from all the bugs with statistical results listed in Table \ref{tab:correlation} and plot the number of patches searched for all successful runs in Figure \ref{fig:searchSpace} together with the trending lines.
The $x$-axis represents the accuracy of fault spaces, and the $y$-axis shows the number of patches searched (in the scale of log2) before a solution is found. 
We can easily observe that GenProg is likely to generate a patch more efficiently with accuracy of the feeding fault space increasing.  

The results shown in Section \ref{RQ1_effectiveness} and Section \ref{RQ1_effeciency} indicate that both the effectiveness and efficiency of GenProg are correlated with accuracy of the feeding fault space.
GenProg is likely to generate more \textit{plausible} and \textit{correct} patches when fault space with high qualities are provided, and it is able to search the solution faster when better fault spaces are provided.
These results deliver the message that fault spaces play an critical role in search-based APR techniques, and the performance of existing techniques is limited by the quality of fault spaces to some extend.
Therefore, besides focusing on designing new repair and searching algorithms, attentions should also be paid on how to design better ways to generate fault spaces with higher quality so as to improve the performance of search-based APR techniques.

\subsection{Adequacy of Test Suite}
\label{sec:coverage}

Sections \ref{sec:faultspace} shows the evidence that $\mathcal{FS}$ highly affects the performance of search-based APR.
Therefore, indicative measurements which approximate the accuracy of fault spaces before applying APR techniques are desired.
In practice, fault spaces are produced by spectrum-based FL techniques when applying APR techniques.
The effectiveness of spectrum-based FL techniques are affected by two factors:
the statistical approaches that calculate the suspicious score \cite{abreu2007accuracy,qi2013using} and the test suite that produces the program spectrum. 
The influence of different statistical approaches to calculate the suspiciousness on search-based APR has already been investigated \cite{assiri2016fault,qi2013using}.
In this study, we focus on how the adequacy of test suite affects APR techniques. 
More specifically, we investigate how different test coverage criteria (e.g., \textit{line, branch} and \textit{mutation}) correlate with the performance of search-based APR, and which coverage criterion correlates the most.
Understanding them is helpful for guiding us in applying APR techniques in practice and designing new tools to generate better fault spaces in the future.
Therefore, we propose the following research question in this part:

\vspace{2mm}

\noindent\fbox{
	\parbox{0.95\linewidth}{
		\textbf{Research Question 2:} How does the fault spaces generated by test suites with different adequacy affect the performance of search-based APR?
	}
}
\vspace{1mm}

Before answering this question, we first describe preparation of test suites with different adequacy and the experimental design.

\subsubsection{Test Case Sampling \& Coverage Measurement}
\label{samplingTest}
In this paper, we use the test coverage to measure the adequacy of test suite \cite{gopinath2014code}.
To control test coverage, we randomly sample subsets of test cases from the original test suite provided by the benchmark dataset. 
Since bugs are exposed by the failing test cases, we keep all the failing test cases.
That is to say, we control the test suite coverage by sampling the passing test cases.
In particular, we follow the heuristic adopted by existing empirical studies to generate different set of test cases \cite{zhang2015assertions}.
We first select the number of intervals $N$ we intent to generate, and the range of each interval is $range=(c_{max}-c_{min}) / N$, where $c_{min}$ is the coverage achieved by the set of failing test cases and $c_{max}$ is the coverage achieved by the original test suite. 
We randomly select test cases from the passing test cases and add them to the targeting sampled test suite together with the failing test cases.
To avoid too many trials of random selection, this process is repeated until the test suite has $(c_{min}+i * range,  c_{min} +　i * range + range]$ when generating a test suite for the $i$-th interval.
We set $N$ to 10 in this study, and generate 20 test suites for each interval.
The sampled test suite is then fed to spectrum based FL technique, and we use the heuristic \textit{Ochiai} to compute the suspiciousness score.
A fault space $\mathcal{FS} = \{\langle l_1, s_1\rangle,\langle l_2, s_2\rangle,...,\langle l_n, s_n\rangle\}$ is produced based on a sampled test suite and then fed to patch generation and validation by APR technique.
In order to isolate the effect of $\mathcal{FS}$, we still use the original test suite for patch validation.
Another reason to adopt the original test suite for validation is that our under-sampling process may induce a weak test suite and it may induce the over-fitting problem \cite{smith2015cure}.
After obtaining the repair results on the 200 runs with different fault space for a bug, we analyze the results and investigate the correlations.

Line, branch and mutation coverages are widely adopted to evaluate the effectiveness and adequacy of a test suite \cite{gopinath2014code,inozemtseva2014coverage}.
We also leverage these three coverage criteria in this study.
We use Cobertula\footnote{http://cobertura.github.io/cobertura/} to measure the line and branch coverage for a test suite, and PiT\footnote{http://pitest.org/} to measure the mutation coverage.
Note that, PiT requires a green test suite for input.
Thus, the existing mutation coverage can only be measured on the sampled passing test cases. 
However, search-based APR techniques generate mutants with heavy bias towards those statements executed by failing test cases.
Motivated by this, we design a new mutation coverage, \textbf{\textit{negative mutation}}, measuring the capabilities of the passing test cases to kill the mutants created on the statements executed by the failing test cases.

We are interested in the correlations between the four coverage criteria and the performance of search-based APR. 
Specifically, we focus on two aspects similar to the previous research question : the effectiveness (e.g., repairing success rate) and efficiency (e.g., the number of patches searched) of search-based APR. 

\subsubsection{Experimental Results}

Similar to RQ1, we focus on those bugs that can be repaired by GenProg within the given time budget during any of the runs.
If GenProg can not produce any solution for all runs with various fault spaces, it may indicate that the correct patch is outside of the search spaces generated by GenProg.
In this case, better algorithms may be desired to produce the search space based on the given fault spaces.
However, this is out of the scope of this study. 

\begin{figure*}[t!]
	\begin{subfigure}{.245\textwidth}
		\includegraphics[width=1.0\linewidth]{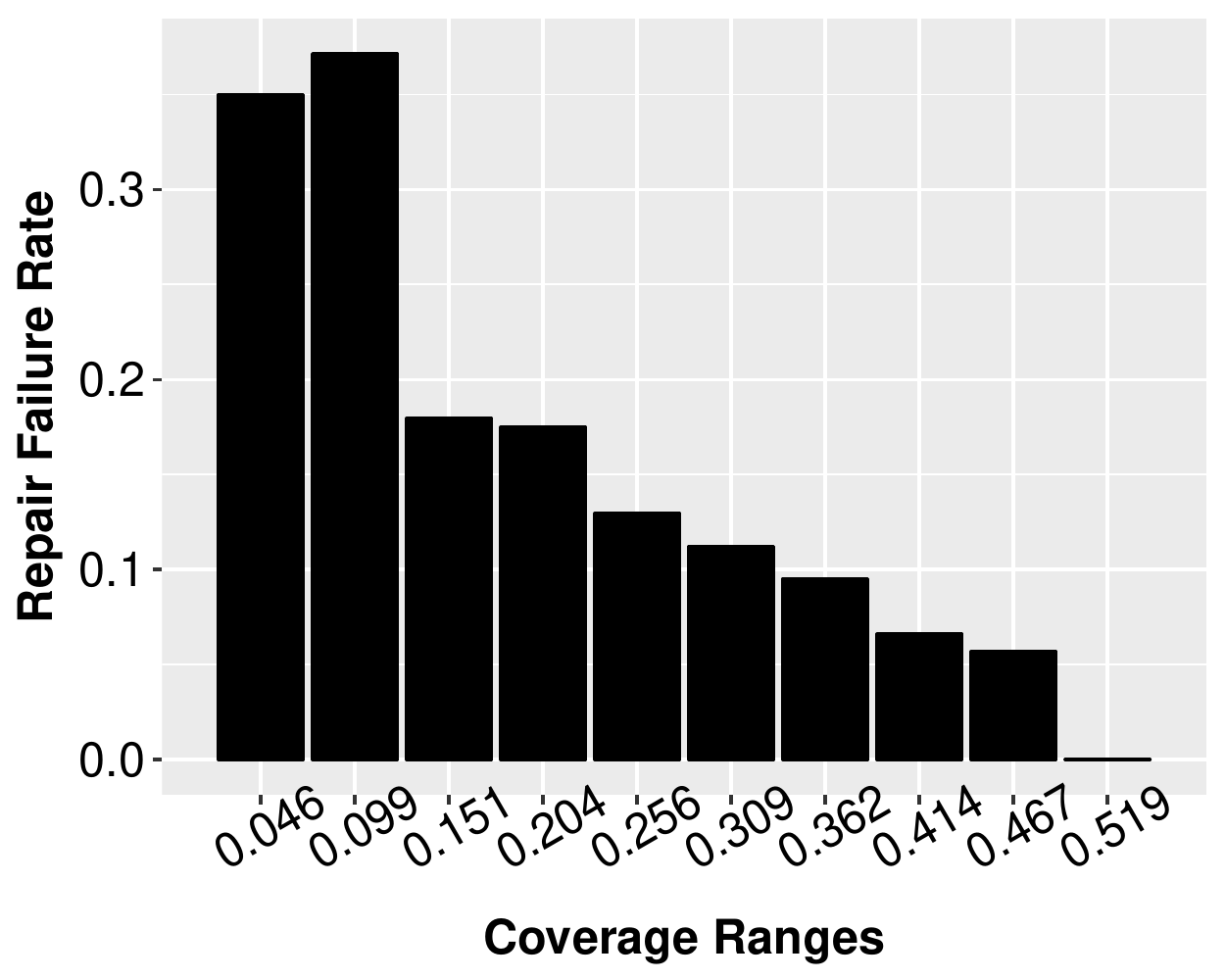}
		\caption{Line}
	\end{subfigure} 
	\begin{subfigure}{.245\textwidth}
		\includegraphics[width=1.0\linewidth]{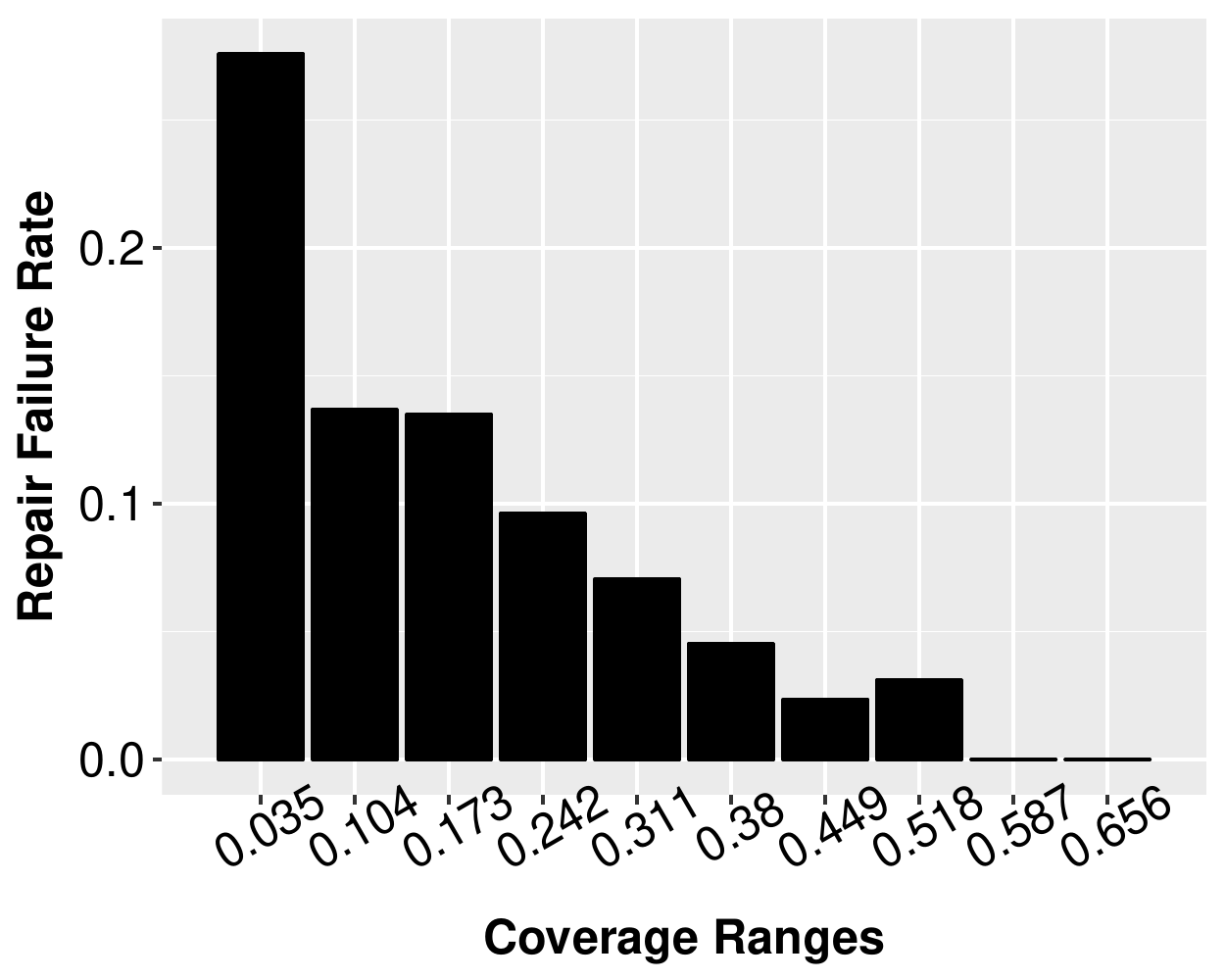}
		\caption{Branch}
	\end{subfigure}
	\begin{subfigure}{.245\textwidth}
		\includegraphics[width=1.0\linewidth]{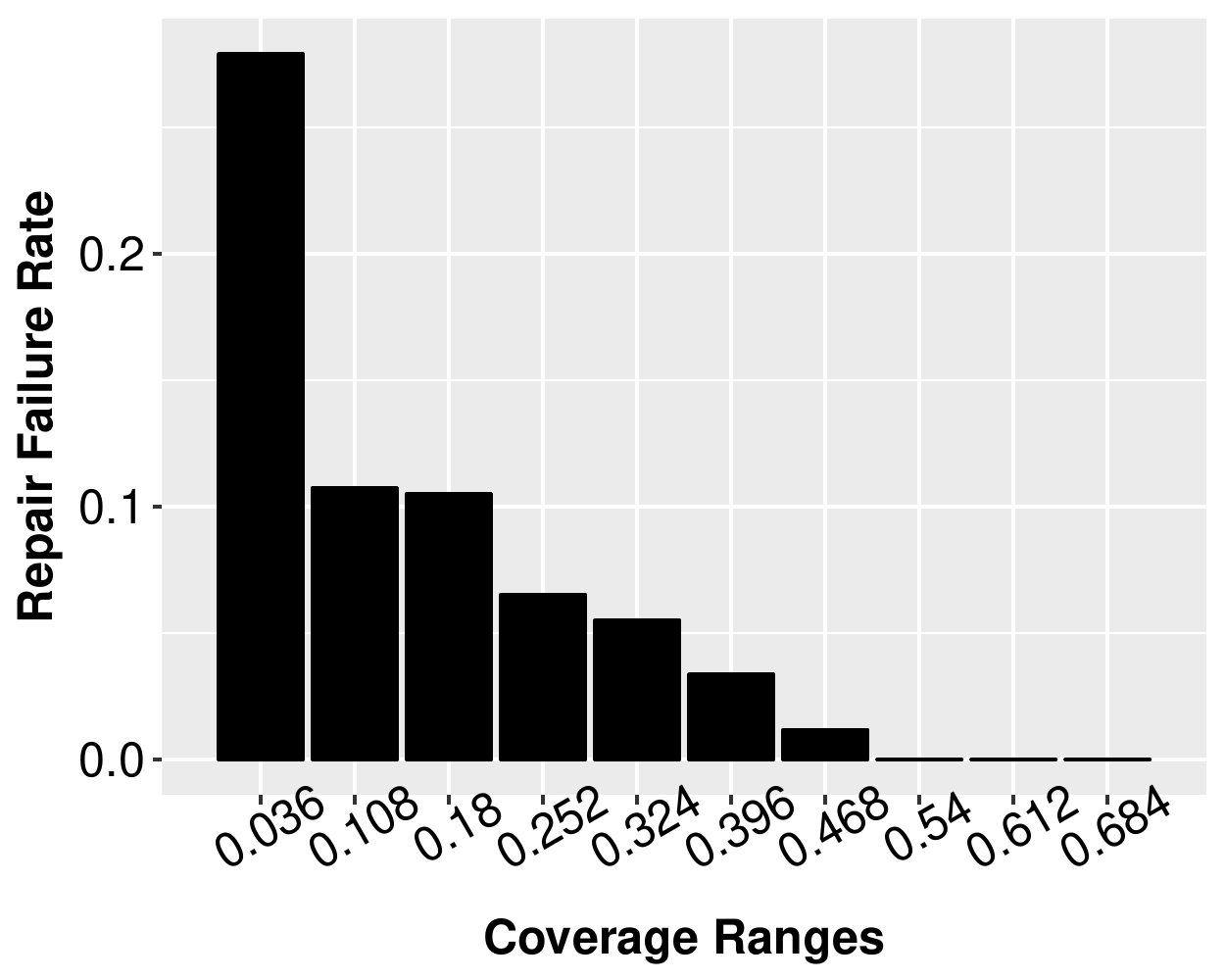}
		\caption{Mutation}
	\end{subfigure}
	\begin{subfigure}{.245\textwidth}
	\includegraphics[width=1.0\linewidth]{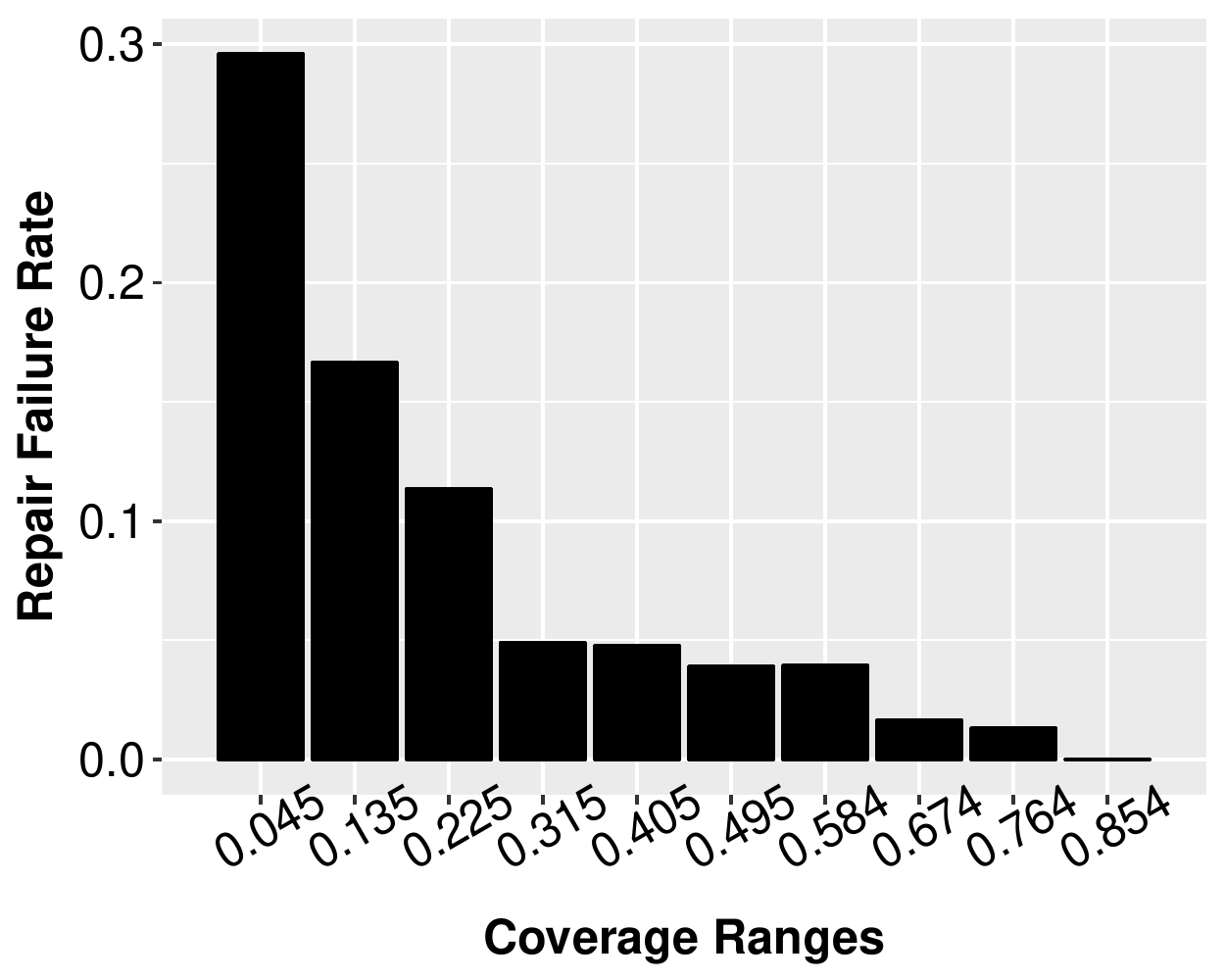}
	\caption{Negative Mutation}
	\end{subfigure} 
	\caption{The Failure Rate of Different Coverage Measures} 
	\label{fig:Coverage_Failing}
	\vspace{-3mm}
\end{figure*}

Figure \ref{fig:Coverage_Failing} shows the failure rate under the 10 ranges of the four different coverage criteria: line, branch, mutation and negative mutation.
The $x$-axis shows the median value for each range.
Note that the coverage ranges for different criteria may be different.
For example, the range of negative mutation coverage is larger than that of the general mutation coverage since we only focus on a specific type of mutants.
For all the four different criteria, the failing rate always decreases with the increasing of coverage ratio in most cases.

\begin{figure*}[t!]
	\centering
	\begin{subfigure}{.2\textwidth}
		\includegraphics[width=1.0\linewidth]{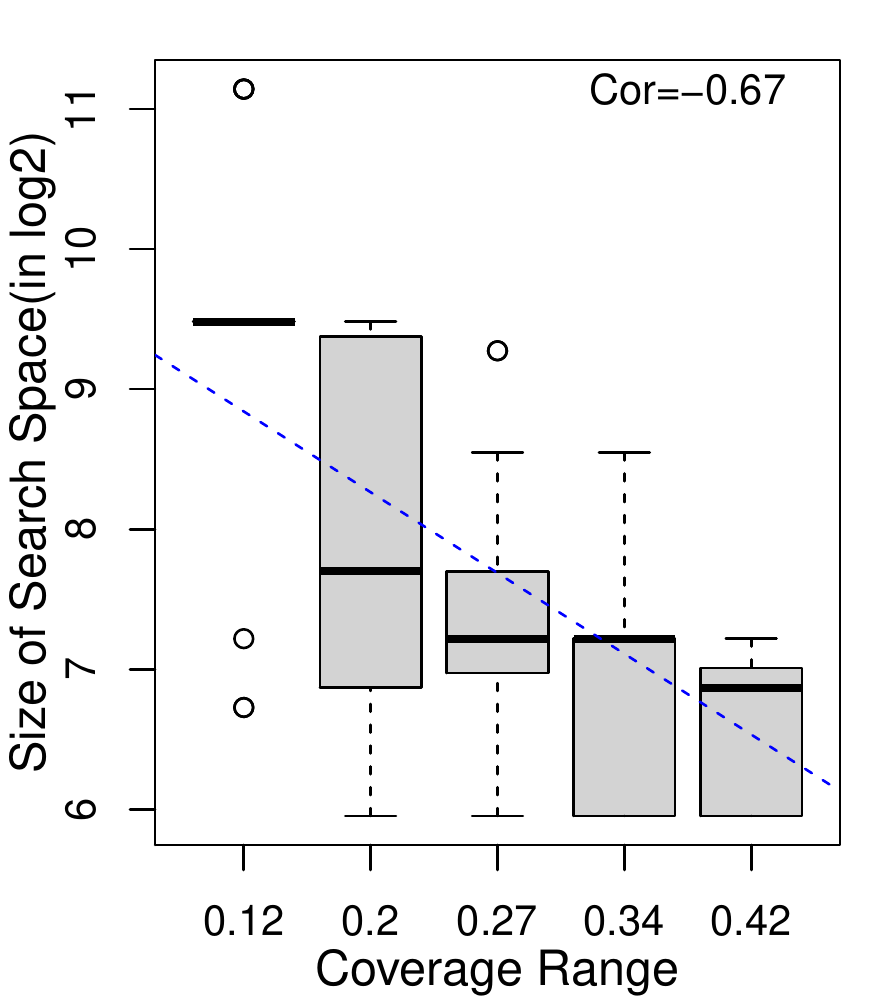}
		\caption{Line}
	\end{subfigure} 
	\begin{subfigure}{.2\textwidth}
		\includegraphics[width=1.0\linewidth]{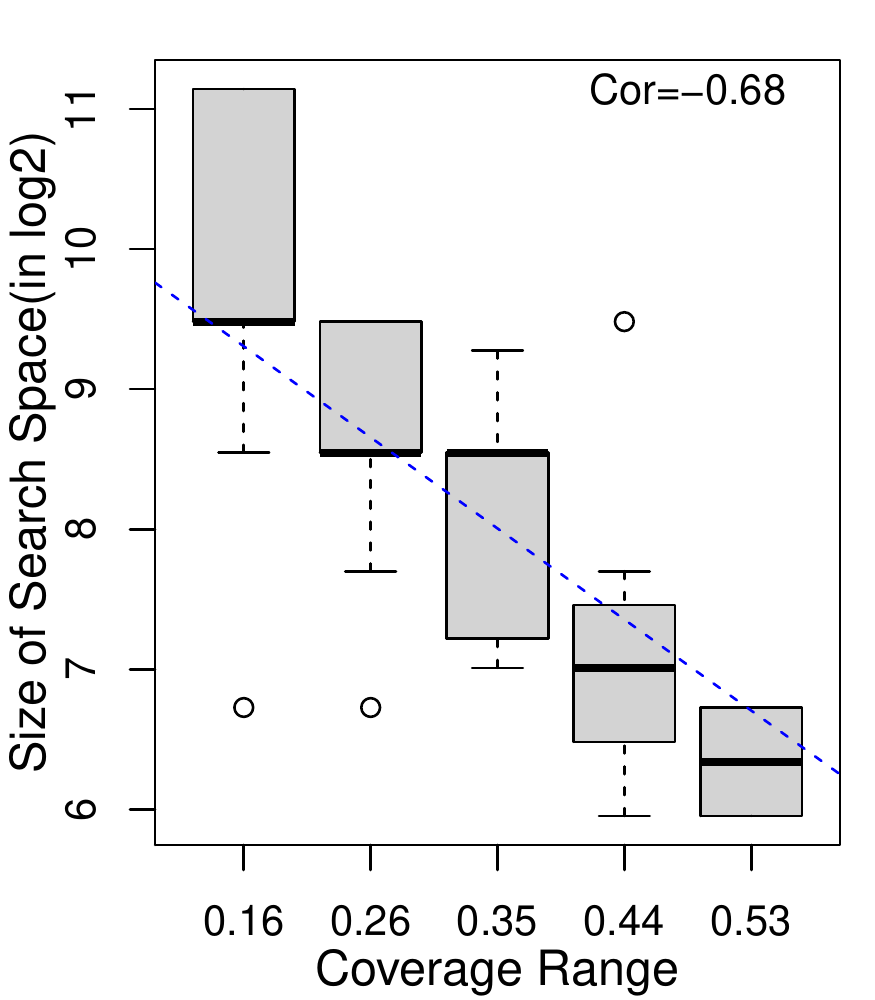}
		\caption{Branch}
	\end{subfigure}
	\begin{subfigure}{.2\textwidth}
		\includegraphics[width=1.0\linewidth]{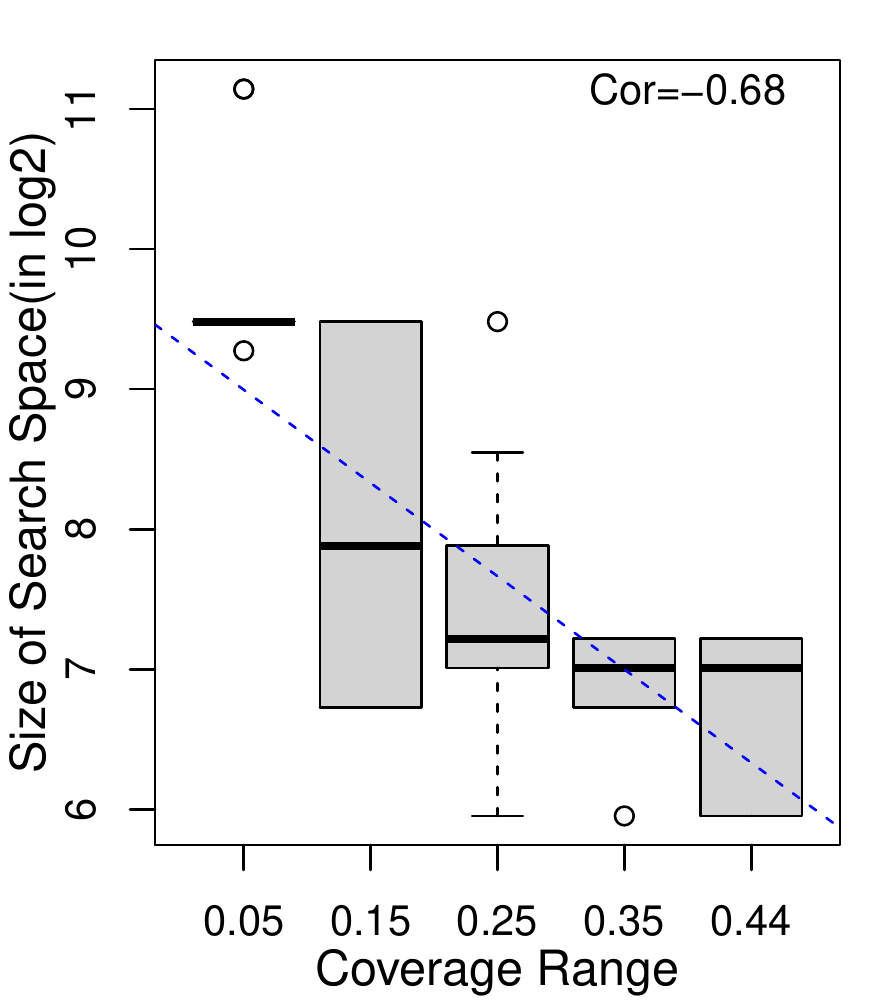}
		\caption{Mutation}
	\end{subfigure}
	\begin{subfigure}{.2\textwidth}
		\includegraphics[width=1.0\linewidth]{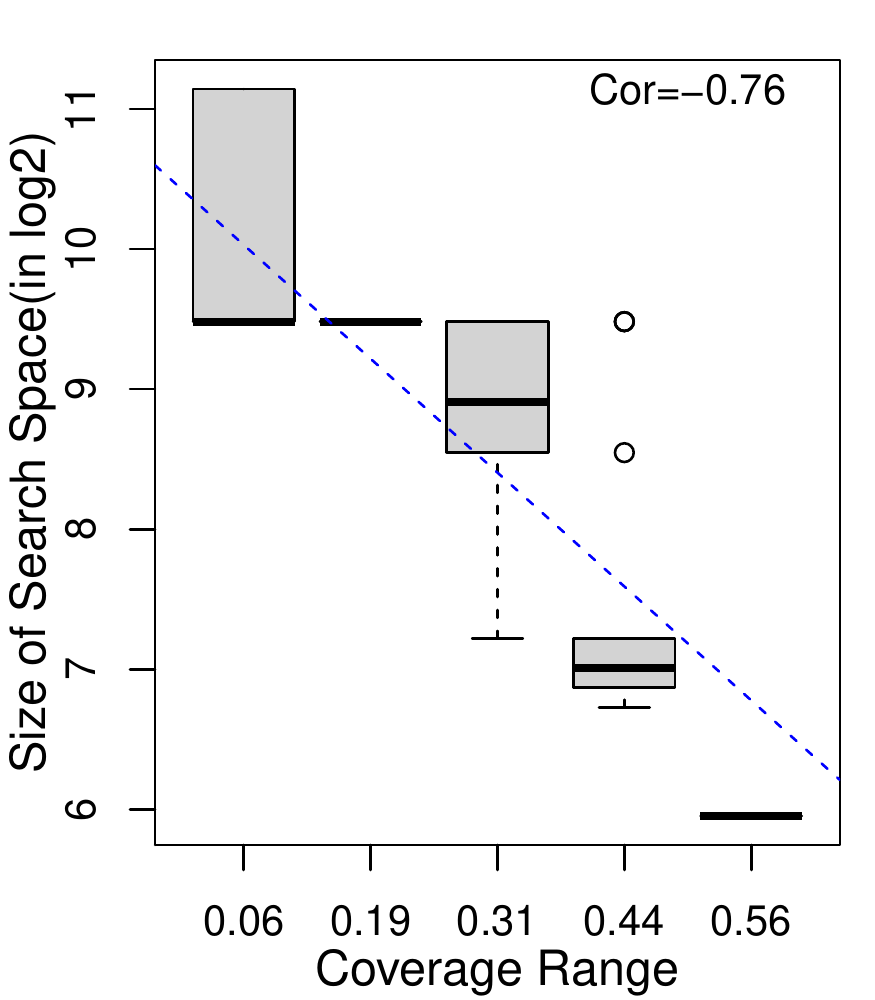}
		\caption{Negative Mutation}
	\end{subfigure} 
	\caption{The Correlations between the Efficiency and Different Coverages of Math 73} 
	\label{fig:Math_73_RQ2}
	\vspace{-3mm}
\end{figure*}

\begin{figure}[h!]
	\centering
	\includegraphics[width=1.0\linewidth]{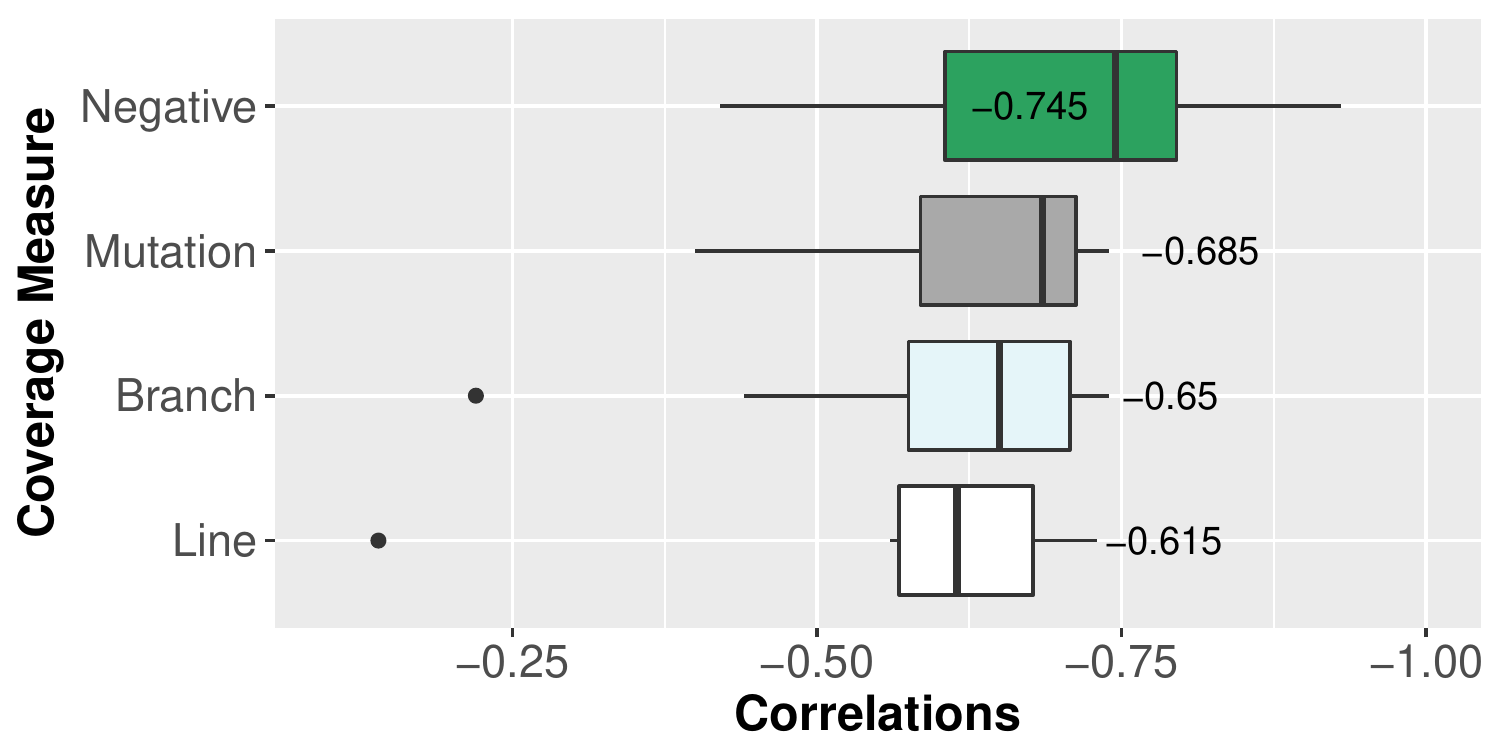}
	\caption{The Correlations between Different Coverage Measures} 
	\label{fig:RQ2}
	\vspace{-5mm}
\end{figure}

For those bugs which have more than 20 successful runs, we analyze the correlations between different coverage criteria and the searching efficiency (e.g., the NPS).
Figure \ref{fig:Math_73_RQ2} plots the results for the four criteria of bug Math 73.
We can observe that all the four coverage criteria are negatively correlated with the number of the patches validated before a solution is found, and the correlation ratio varies from $-0.67$ to $-0.76$.
These results indicate that when we improve the coverage of the test suite to generate fault spaces, search-based APR technique is able to find the solution faster if the solution is in the search space.
Similar results can be also be found for the other bugs, and Figure \ref{fig:RQ2} plots the correlation values for all bugs with statistical results. 
We can observe that among all the four coverage criteria, the negative mutation score has the strongest correlation in general.
The Mann-Whitney U-Test was conducted to see if the correlation measured by negative mutation score is significantly smaller than those of the other measurements.
The maximum p value is $0.0438$, which confirms that the negative mutation score is significantly better than the other metrics when measuring the correlations between coverages and the performance of search-based APR.


These results reveal that fault spaces generated by different test suites affect the performance of search-based APR techniques.
They also deliver valuable messages to guide us when applying search-based APR techniques in practice.
First, when generating fault spaces, test suites with higher coverage are always desired, since it may help increase the repairing success rate and reduce the number of patches to be validated.
Such improvement can bring a lot of savings, since validating a patch by running it against the whole test suite has been demonstrated to be extremely expensive \cite{qi2014strength,weimer2013leveraging}.
Second, the negative mutation coverage has the strongest correlation with the searching efficiency. 
This indicates that, by improving the negative mutation score, we have higher probabilities to find the solution faster.

\subsection{Automatically Generated Test Cases}
\label{sec:automatedTest}

Automated test generation (ATG) tools like Randoop \cite{pacheco2007randoop} and Evosuite \cite{fraser2013whole} are capable of generating unit tests automatically.
These automatically generated tests are capable of increasing the coverage of existing test cases, and they are also found to be capable of detecting real faults \cite{shamshiri2015automatically}.
Motivated by this and the findings drawn from our RQ2, we are interested in the capabilities of these automatically generated tests in improving the performance of search-based APR.
Therefore, we proposed the following research question:

\vspace{2mm}
\noindent\fbox{
	\parbox{0.95\linewidth}{
		\textbf{Research Question 3:} Can the fault spaces generated by test suites augmented with ATG tools help increase the performance of search-based APR?
	}
}
\vspace{2mm}

\subsubsection{Test Suite Augmentation}

We adopt Randoop \cite{pacheco2007randoop} and Evosuite \cite{fraser2011evosuite} to augment the original test suite provided by the benchmark for each bug in this study.
Randoop is a feedback-directed random test generation tool for object-oriented programs \cite{pacheco2007randoop}.
It extends the call sequences iteratively by randomly selecting candidates until the generated sequence raises an unexpected behavior or violates any code contract.
Evosuite leverages a genetic algorithm in order to evolve a set of test cases which maximizes code coverage \cite{fraser2011evosuite}.
The settings of generating test cases we use are similar to existing empirical studies \cite{shamshiri2015automatically}.
For Randoop, it requires a list of target classes as input to explore during test generation since it adopts a bottom-up approach which requires all dependencies of a method call. 
Here we select all the classes in the targeted package as its input,
and run it three minutes for each buggy version by using different random seeds.
For Evosuite, we run it for one minute to generate tests using different random seeds for each class under the targeted package, since this tool targets a specific class under test.
For the other options, we use the default settings.

Actually, not all of test cases generated on the \textit{buggy version} of the program are used in our study. 
We remove those invalid test cases (e.g., those can not be compiled). 
We also remove the test cases that fail on the \textit{fixed version} of the program since these failing test cases are not caused by the bugs we studied.
This is because, if we keep those failing test cases irrelevant to the bugs we studied, it may produce bias influence for our conclusion. 
Based on the generated test cases after filtering and the original test suite, we then generate the corresponding fault spaces.
In this part, the spectrum-based approach used to compute the suspicious score is also \textit{Ochiai}.
Finally, we record the results for each run by feeding the generated fault spaces to GenProg.


\subsubsection{Experimental Results}

In this research question, our interests focus on whether the augmented test suites can improve the accuracy of fault spaces, and thus improve the performance of search-based APR in a further step.
Table \ref{tab:RQ3} shows the highlights of the results.
We record the ATG tool and the seed used to generate the extra tests. 
``\#P'' indicates the number of valid passing test cases generated, and ``\#F'' indicates the number of valid failing test cases generated.
We observe that generating valid failing tests which reveal the targeted bug is difficult for ATG tools.
Plenty of valid passing test cases are generated, the number of which varies from less than 100 to more than one thousand.
Starting from the 7$^{th}$ column (the ``MAP'' column), there are two values in each cell.
The first value shows the information of the original test suite while the second indicates the information of the test suite after test augmentation.
All the automatically generated test cases can improve the coverages for all the four different criteria.

\begin{table*} [h]
	\centering
	\scriptsize
	\caption{Repairing Results Using Fault Space Generated with Augmented Test Cases}
	\label{tab:RQ3}
	\bgroup
	\setlength\tabcolsep{2.6pt}     
	\def\arraystretch{1.2}
\begin{tabular}{ll|cccc|cl|cl|cc|cc|cc|cc|cl} \hline
	Project  & Id & ATG & Seed & \#P & \#F & \multicolumn{2}{c|}{MAP} &  \multicolumn{2}{c|}{MRR} & \multicolumn{2}{c|}{Line} & \multicolumn{2}{c|}{Branch} & \multicolumn{2}{c|}{Mutation} & \multicolumn{2}{c|}{Negative} & \multicolumn{2}{c}{NPS}    \\ \hline
	
	\multirow{2}{*}{Math} & \multirow{2}{*}{5$^\surd$}   & Randoop    &  762    &  653  &   0  &   0.500  &  1.000$^{\uparrow}$   & 0.500  &  1.000$^{\uparrow}$  & 0.577   &  0.623$^{\uparrow}$  &  0.758  &  0.819$^{\uparrow}$ &  0.800   &  0.862$^{\uparrow}$   &   0.723  &  0.862$^{\uparrow}$  &  306 &   28$^{\uparrow}$  \\ 
	 &    & Evosuite    &  627    &  55  &   0  &   0.500  &  1.000$^{\uparrow}$   & 0.500  &  1.000$^{\uparrow}$  &  0.577   &  0.623$^{\uparrow}$   &   0.758  &  0.824$^{\uparrow}$  & 0.800   &  0.888$^{\uparrow}$  &  0.723  & 0.862$^{\uparrow}$   & 306 &   28$^{\uparrow}$  \\ \hline
	
	\multirow{2}{*}{Math} & \multirow{2}{*}{50$^\surd$} & Randoop    &  734    &  9155  &   0  &   0.917  &  0.917$^{\rightarrow}$   & 1.000  &  1.000$^{\rightarrow}$  &  0.594   &  0.608$^{\uparrow}$   &   0.553  &  0.557$^{\uparrow}$  & 0.645   &  0.652$^{\uparrow}$  &  0.833  & 0.856$^{\uparrow}$  & 46 &   46$^{\rightarrow}$  \\
	 &    & Evosuite    &  974    &  215  &   0  &   0.917  &  0.917$^{\rightarrow}$   & 1.000  &  1.000$^{\rightarrow}$  &  0.594   &  0.615$^{\uparrow}$  &   0.553  &  0.572$^{\uparrow}$  & 0.645   &  0.699$^{\uparrow}$  &  0.833  &  0.889$^{\uparrow}$  & 46 &   46$^{\rightarrow}$  \\ \hline
	
	\multirow{2}{*}{Math} & \multirow{2}{*}{53$^\surd$}  & Randoop    &  282    &  516  &   0  &   1.000  &  0.500$^{\downarrow}$   & 1.000  &  0.500$^{\downarrow}$  &  0.605   &  0.658$^{\uparrow}$   &  0.774  &  0.840$^{\uparrow}$  & 0.799   &  0.862$^{\uparrow}$  &  0.897  & 0.923$^{\uparrow}$  & 100 &   403$^{\downarrow}$  \\
	 &   & Evosuite    &  416    &  79  &   0  &   1.000  &  1.000$^{\rightarrow}$   & 1.000  &  1.000$^{\rightarrow}$  & 0.605   &  0.667$^{\uparrow}$   &   0.774  &  0.840$^{\uparrow}$  & 0.799   &  0.890$^{\uparrow}$  &  0.897  &  0.949$^{\uparrow}$  & 100 &   100$^{\rightarrow}$  \\ \hline
	
	\multirow{2}{*}{Math} & \multirow{2}{*}{70$^\surd$} & Randoop    &  667    &  3141  &   0  &   1.000  &  1.000$^{\rightarrow}$   & 1.000  &  1.000$^{\rightarrow}$  &  0.550   &  0.573$^{\uparrow}$   &   0.646  &  0.650$^{\uparrow}$  & 0.557   &  0.617$^{\uparrow}$  &  0.421  & 0.658$^{\uparrow}$  & 62 &   62$^{\rightarrow}$  \\
	&    & Evosuite    &  582    &  83  &   0  &   1.000  &  1.000$^{\rightarrow}$   & 1.000  &  1.000$^{\rightarrow}$  &  0.550   &  0.575$^{\uparrow}$  &   0.646  &  0.694$^{\uparrow}$  & 0.557   &  0.641$^{\uparrow}$  &  0.421  &  0.500$^{\uparrow}$  & 62 &   62$^{\rightarrow}$  \\ \hline
	
	\multirow{2}{*}{Math} & \multirow{2}{*}{73$^\surd$} & Randoop    &  793    &  4518  &   0  &   1.000  &  1.000$^{\rightarrow}$   & 1.000  &  1.000$^{\rightarrow}$  &  0.547   &  0.571$^{\uparrow}$   &   0.637  &  0.642$^{\uparrow}$  & 0.558   &  0.619$^{\uparrow}$  &  0.650  & 0.671$^{\uparrow}$  & 62 &   62$^{\rightarrow}$  \\
	&    & Evosuite    &  167    &  84  &   0  &   1.000  &  1.000$^{\rightarrow}$   & 1.000  &  1.000$^{\rightarrow}$  &  0.547   &  0.561$^{\uparrow}$  &   0.637  &  0.652$^{\uparrow}$  & 0.558   &  0.605$^{\uparrow}$  &  0.650  &  0.675$^{\uparrow}$  & 62 &   62$^{\rightarrow}$  \\ \hline

	\multirow{2}{*}{Chart} & \multirow{2}{*}{1$^\surd$}   & Randoop    &  95    &  1641  &   0  &   0.125  &  0.333$^{\uparrow}$   & 0.125  &  0.333$^{\uparrow}$  &  0.446  &  0.460$^{\uparrow}$   &   0.432  &  0.465$^{\uparrow}$  & 0.078   &  0.144$^{\uparrow}$  &  0.449  &  0.816$^{\uparrow}$  & - & 56$^{\uparrow}$  \\ 
	& & Evosuite    &  390    &  580  &   0  &   0.125  &  0.250$^{\uparrow}$   & 0.125  &  0.250$^{\uparrow}$  &  0.446  &  0.476$^{\uparrow}$   &   0.432  &  0.503$^{\uparrow}$  & 0.078   &  0.138$^{\uparrow}$  &  0.449  &  0.694$^{\uparrow}$  & - & -  \\ \hline

	\multirow{2}{*}{Math} & \multirow{2}{*}{85$^\star$} & Randoop    &   308   &  2008  &   0  &   0.091  &  0.091$^{\rightarrow}$   & 0.091  &  0.091$^{\rightarrow}$  &  0.541   &  0.567$^{\uparrow}$   &   0.652  &  0.667$^{\uparrow}$  & 0.556   &  0.617$^{\uparrow}$  &  0.444  & 0.444$^{\uparrow}$  & 53 &   53$^{\rightarrow}$  \\
	&    & Evosuite    &  619    &  81  &   0  &   0.091  &  0.167$^{\uparrow}$   & 0.091  &  0.167$^{\uparrow}$  &  0.541   &  0.550$^{\uparrow}$  &   0.652  &  0.676$^{\uparrow}$  & 0.556   &  0.619$^{\uparrow}$  &  0.444  &  0.617$^{\uparrow}$  & 53 &   53$^{\rightarrow}$  \\ \hline
	
	\multirow{2}{*}{Math} & \multirow{2}{*}{95$^\star$}   & Randoop    &  413    &  1458  &   0  &   0.417  &  0.417$^{\rightarrow}$   & 0.333  &  0.333$^{\rightarrow}$  &  0.521   &  0.526$^{\uparrow}$   &   0.720  & 0.777$^{\uparrow}$  &  0.734  & 0.869$^{\uparrow}$   &  0.851  &  0.946$^{\uparrow}$ & 814 &   814$^{\rightarrow}$  \\ 
	&   & Evosuite    &  492    &  455  &   0  &   0.417  &  0.583$^{\uparrow}$   & 0.333  &  0.500$^{\uparrow}$  &  0.521   &  0.589$^{\uparrow}$   &   0.720  & 0.767$^{\uparrow}$  &  0.734  & 0.851$^{\uparrow}$   &  0.851  &  0.946$^{\uparrow}$ & 814 &   736$^{\uparrow}$  \\ \hline
	
	\multicolumn{20}{l}{\tiny{``$^\surd$'' indicates that correct patches are generated for this bug. ``$^\star$'' indicates that plausible patches are generated for this bug.}} \\ 
	\multicolumn{20}{l}{\tiny{NPS indicates the number of patches searched until a solution is found.``-'' indicates that no solution is generated within one hour.}}
	 \\
\end{tabular}
	\egroup
\vspace{-5mm}	
\end{table*}

The results in Table \ref{tab:RQ3} list all the bugs can be correctly repaired, and they reveal that automatically generated unit tests can actually help improve the performance of search-based ARP techniques for some cases.
For example, for Chart 1, GenProg is incapable of generating a patch given the fault space produced by the original test suite within one hour.
After test augmentation, the negative mutation coverage has been increased to $82\%$. 
The accuracy of the generated fault space has also been improved remarkably.
Feeding this fault space, GenProg is capable of finding the correct patch after searching for 56 candidate patches.
For Math 5, the efficiency of GenProg can also been improved by augmenting the original test suite.
We show the original fault space and the fault spaces generated using ATG tools in Table \ref{tab:RQ3_MATH5}.
Originally, the buggy statement (line 305) is ranked the same as another clean statement (line 1228), and the minimum number of the searched patches for GenProg is 306 (Table \ref{tab:RQ3}).
By augmenting the test suite using Randoop with seed 762 or Evosuite with seed 627, the tie of the rank can be broken by picking line 305 in prior of the others.
With more accurate fault space, it only takes GenProg 28 validations before finding the correct patch.
Similar case can also be found for Math 95 as shown in Table \ref{tab:RQ3}.

We also observe some cases whose fault space remains unchanged after the test argumentation, such as Math 50, Math 70, Math 73, and Math 85 as shown in Table \ref{tab:RQ3}.
Accordingly, the performance of GenProg does not change among these cases.
For Math 50, Math 70, Math 73, the accuracy of the fault spaces produced by original test suites is very high (their buggy statements are ranked top 1). 
Therefore, it is not surprising that GenProg cannot be improved, since the effect of augmented test cases is limited. 
For Math 85, although the augmented test cases generated by Evosuite improve the accuracy of the fault space, the performance of GenProg does not change, since it generates the plausible patch in a clean statement. 
There is also an exceptional case, Math 53, that the augmented test cases decrease the accuracy of the fault space.
It includes only 8 statements executed in the failing test case, and buggy statements have been executed frequently even in passing test cases.  
 
In summary, these results confirm that test cases generated by ATG tools can help improve the performance of search-based APR.


\begin{table} [h]
	\centering
	\small
	\caption{Different Fault Spaces of Math 5}
	\label{tab:RQ3_MATH5}
	\bgroup
	\setlength\tabcolsep{6pt}     
	\def\arraystretch{1.1}
	\begin{tabular}{r|c|c|c}
		\hline
		& Original & Randoop 762 & Evosuite 627 \\ \hline
		Top Elements & 1228 : 1.000 &  305 : 0.447     &  305 : 0.707     \\ 
		(\textit{line: susp})& 305 : 1.000 & 304 : 0.174      &  217 : 0.316     \\
		\multirow{3}{*}{}	& 304 : 0.447 &  1228 : 0.164     &    204 : 0.316   \\
		                    & 300 : 0.408    &  348 : 0.154     &  130 : 0.289     \\ \hline
		\multicolumn{4}{l}{\tiny{All the lines come from class \textbf{Complex.java}}, and line 305 is the oracle}  \\
	\end{tabular}
	\egroup
\vspace{-6mm}	
\end{table}

\subsection{Threats to Validity}

Our work suffers from several threats to validities.
First, the experiments may not generalize.
The controlled experiments are only conducted on GenProg, one of the most representative search-based APR technique.
The results may not extend to other search-based techniques, which leverage different operators to create mutants.
However, recent work has started to unify the theory underlying the search-based APR techniques \cite{weimer2013leveraging}, which suggests that results from two different techniques may extend to others.
Besides, this study focus on the influence of fault spaces on search-based APR.
Currently, the methods to produce fault spaces adopted by existing techniques are pretty much the same, and thus they face similar challenges in terms of the quality of fault spaces.
For example, if the accuracy of a fault space is low or even do not contain the real fix location, search-based APR is unable to generate a patch no matter what operators are leveraged to generate mutants.
Therefore, some of our findings could still deliver valuable information for other techniques.

Second, the ways to measure the accuracy of fault spaces may not be perfect.
In this study, we leverage the fixing patch submitted by developer as the oracle to measure the accuracy of a fault space.
However, existing studies have reported that there might be multiple ways to correctly fix a bug \cite{murphy2013design,murphy2015design}, and developers may submit a work-around solution to resolve the issue instead of fixing the bug essentially.
Therefore, only considering the locations changed in the fixing patch may not always be correct.
In the future, we plan to conduct program analysis based on the fixing patch in order to prepare more accurate fixing oracles.

Third, we focus on Java bugs in this study, which incur the thread to validity that our conclusion may not be generalizable to other program like C/C++.
Besides, the dataset considered may not be representative to all types of bugs.
However, Defects4J was proposed in order create a large, unbiased, and realistic benchmark set \cite{just2014defects4j}.
It has been widely adopted and evaluated by existing studies recently \cite{martinez2016automatic,shamshiri2015automatically,xuan2016nopol,le2016history,martinez2016astor}, which help mitigate this issue.

%% file: related.tex
\section{Related Work}

\subsection{Automated Program Repair}
Automated program repair techniques can be broadly classified into two categories.
The first category is seach-based APR. besides the most representative one, GenProg \cite{le2012systematic,weimer2009automatically}, there are also some other techniques.
RSRepair \cite{qi2014strength} adopts the same fault localization technique and mutation operators as GenProg.
It differs from GenProg in the searching algorithm leveraged.
Instead of using genetic programming, it searches among all the candidates randomly.
The dominant cost of search-based APR techniques comes from the process of validating candidate patches by running test cases \cite{forrest2009genetic}.
Therefore, reducing the number of candidate patches and reducing the test cases required for validating a patch are two important aspects to boost the efficiency of search-based APR techniques.
Motivated by this, AE \cite{weimer2013leveraging} proposes two heuristics \textit{RepairStrat}, which leverages program equivalence relations to prune semantically-equivalent patches, and \textit{TestStrat}, which samples test cases for validation, to reduce the total cost required to find a correct patch.
In order to generate candidates that are more likely to be the correct patch, PAR \cite{kim2013automatic} proposes to leverage fixing templates learned from human patches to generate possible candidates.
Each template is able to fix a common type of bugs.
Furthermore, Debroy \cite{debroy2010using} proposes to use standard mutations from the mutation testing literature to fix programs.
Relifix \cite{tan2015relifix} targets at fixing regression bugs.
The second category is semantics-based APR.
Staged Program Repair (SPR) \cite{long2015staged} is a hybrid of search-based and semantics-based automated program repair technique.
It leverages traditional mutation operators to generate candidate patches, and it is also capable of synthesizing conditions via symbolic execution.
Prophet \cite{long2016automatic} was proposed based on SPR, which is capable of prioritizing candidate patches via learning from correct patches automatically using machine learning techniques.
SPR and Prophet only synthesize program elements involving conditions.
SemFix \cite{nguyen2013semfix} is capable of synthesizing right hand side of assignments besides statements involving conditions.
Angelix \cite{mechtaev2016angelix} was proposed to address the scalability issue concerning semantics-based techniques.
The key enabler for scalability is the novel lightweight repair constraint proposed in this paper, which is called \textit{angelic forest}, which is automatically extracted via symbolic execution.
Other semantics-based APR techniques were also proposed and well evaluated \cite{mechtaev2015directfix,demarco2014automatic,xuan2016nopol}.
Both search-based APR and semantics-based APR techniques have their advantages and disadvantages.
Search-based APR is simple, intuitive and is generalizable to fix all types of bugs, which render them more effective.
However, the efficiency is greatly compromised by the search space explosion problem \cite{long2016analysis}. 
On the other hand, semantics-based APR is more effective since the search space is more tractable by using program synthesis with restricted components.
However, the effectiveness could be limited by the capability of constraint solving and program synthesis.
Besides, search-based APR is scalable to fix large-scale programs which semantics-based APR is less scalable due to the overhead of symbolic execution and constraint solving.

\subsection{Empirical Evaluations of APR}
There are plenty of existing empirical studies related to APR techniques.
One work that closely relates to this study is conducted by Qi \all \cite{qi2013using}, which empirically evaluated the effectiveness of fault localization techniques in the domain of automated program repair.
Qi \all compared the effectiveness of 15 different fault localization techniques using GenProg.
Different FL techniques may generate different fault spaces, and affect the effectiveness the GenProg.
They also use the number of patches searched to measure the effectiveness.
An parallel study \cite{assiri2016fault} also investigated the role of fault localization in automated program repair.
Different to them, we evaluate the influence of fault spaces on search-based APR directly.
Specifically, how dose the accuracy of fault spaces correlate with the performance of search-based APR.
Despite the FL techniques, the accuracy of fault spaces can also be influenced by other factors, such as the leveraged test suite, and we also investigate the correlation between the adequency of test suites and the performance of search-based APR in this study.
Another work relates to this study is the one which evaluates the overfitting problem in search-based APR \cite{smith2015cure}.
The authors also sampled the test cases in order to prepare different test suites with different coverage or different number of failing test cases.
They used the sampled test suite in validation in order to investigate whether patches validated with fewer test cases are more likely to overfit. 
However, in our study, the sampled test cases are only leveraged to generate the fault spaces since we control the fault space as the sole variable in this study.
In such a way, we control the quality of fault spaces in order to investigate its correlation with the performance of search-based APR.
Long \all analyzed the search spaces for search-based APR techniques systematically \cite{long2015staged}.
It focused on the distributions of the correct patches among all the candidates, and found that the correct patches are sparse in the search spaces.
It also characterized the trade-off in the structure of the search spaces.
Other empirical studies investigated the maintainability of the patches generated by APR techniques \cite{fry2012human}.
Monperrus \cite{monperrus2014critical} further discussed the patch acceptability criteria of machine-generated patches and emphasized that it requires a high level of expertise to evaluate the patch acceptability. 
Tao \all proposed to leverage machine-generated patches as debugging aid, which is used to facilitate real developers in understanding the bug and fixing it with less effort \cite{tao2014automatically}, and their experimental results shed light on other applications and usefulness of the patches generated by APR techniques. 

%% file: conclusion.tex
\section{Conclusion}

Search-based APR techniques are one of the important directions for automatic program repair \cite{le2012systematic,weimer2009automatically,weimer2013leveraging,kim2013automatic,long2015staged,long2016automatic}.
The effectiveness of these techniques is greatly compromised by the search space explosion problem.
One key dimension defines the search space is the fault space.
Therefore, understanding the influence of fault space on search-based APR techniques is important.
In this paper, we conducted an empirical study on understanding this influence. 
We observe that the accuracy of fault space significantly affects the effectiveness and efficiency of search-based APR techniques.
We also find that test coverages correlate significantly with the performance of search-based APR techniques.
Among different coverage measurements, the negative mutation coverage, which measures the capability of a test suite to kill the mutants created on those statements executed by failing test cases, is the most indicative measurement to estimate the accuracy of the fault space.
Moreover, our results confirm that argument test suites generated by automated test generation tools have the potential to improve the performance of search-based APR.
These results motivate us to generate better fault spaces automatically when designing APR techniques in the future.

